\documentclass{pegasus}

\allowdisplaybreaks
\begin{document}

\title{Going beyond ER=EPR in the SYK model}

\author[a]{Micha Berkooz,}
\author[a]{Nadav Brukner,}
\author[b]{Simon F. Ross,}
\author[a]{and Masataka Watanabe}
\emailAdd{micha.berkooz@weizmann.ac.il}
\emailAdd{nadav.brukner@weizmann.ac.il}
\emailAdd{s.f.ross@durham.ac.uk}
\emailAdd{masataka.watanabe@weizmann.ac.il}
\affiliation[a]{Department of Particle Physics and Astrophysics, \\ Weizmann Institute of Science, Rehovot 7610001, Israel}
\affiliation[b]{Centre for Particle Theory, Department of Mathematical Sciences \\
Durham University, Stockton Road, Durham DH1 3LE, UK}

% \texorpdfstring{${}^a$}{}
\abstract{
We discuss generalizations of the TFD to a density matrix on the doubled Hilbert space. We suggest that a semiclassical wormhole corresponds to a certain class of such density matrices, and specify how they are constructed. Different semi-classical profiles correspond to different non-overlapping density matrices. We show that this language allows for a finer criteria for when the wormhole is semiclassical, which goes beyond entanglement. Our main tool is the SYK model. 
We focus on the simplest class of such density matrices, in a scaling limit where  the ER bridge is captured by chords going from one space to another, encoding correlations in the microscopic Hamiltonian. The length of the wormhole simply encodes the extent these correlations are eroded when flowing from one side to the other.
}

\maketitle

%%%%%%%%%%%%%%%%%%%%%%%%%%%%%%%%%%%%%%%%%

\newpage

\section{Introduction}

A key development in the holographic study of black holes was Maldacena's identification of the thermofield double state with the eternal black hole \cite{Maldacena:2001kr}. The thermofield double state (TFD) is an entangled state in two copies $\mathcal H_L \otimes \mathcal H_R$ of a quantum system, 
\begin{align} \label{eq:TFD_Def}
    \begin{split}
    \left|\text{TFD\ensuremath{_{\beta}}}\right> & =Z^{-1/2}\left(\beta\right)\sum_{n}e^{-\beta E_{n}/2}\left|n\right>_{L}\left|n\right>_{R}.
    \end{split}
\end{align}
 This state can be prepared by a Euclidean path integral over an interval of length $\beta/2$ in Euclidean time. In a holographic system, an uncharged black hole provides a bulk saddle-point for these Euclidean boundary conditions; in situations where this is the dominant saddle-point, on continuation to the Lorentzian picture the eternal maximally extended black hole then provides the dual description of the state \eqref{eq:TFD_Def}. This connection is a key motivating example for the idea that entanglement is related to connection of the bulk geometry \cite{Maldacena:2013xja}. There are two-point functions involving operators in the two copies of the system $\langle \mathcal O_L \mathcal O_R \rangle$ which have non-zero values in this state; in the connected bulk geometry, influences can propagate from one boundary to the other, while in the boundary quantum theory, the two-point functions can be non-vanishing in the absence of any interaction due to the entanglement of the state \cite{Garcia-Garcia:2020ttf}. 

We would like to know more generally what states in $\mathcal H_L \otimes \mathcal H_R$ can have such a wormhole description. Entanglement is clearly necessary (in its absence, correlations $\langle \mathcal O_L \mathcal O_R \rangle$ in non-interacting systems would vanish), but it has been argued that it is not sufficient. In \cite{Marolf:2013dba,Balasubramanian:2014gla}, it was argued that in a generic state with the same degree of entanglement as the TFD, the correlations of simple operators between the two boundaries is exponentially small, so generic entangled states do not have a geometric wormhole description. Some studies have considered small departures from the TFD, finding that these lead to longer wormholes. This was first explored in \cite{Shenker:2013pqa}, who considered a local operator insertion on the TFD evolved in Lorentzian time. A related investigation in \cite{Goel:2018ubv} considered operator insertions in the Euclidean path integral defining the TFD state. We will review these studies in the next section. 

Our aim in this paper is to further explore when states in two copies of a holographic system have a bulk wormhole description. We will argue that as we move away from the TFD, it is natural to consider {\it mixed states on} $\mathcal H_L \otimes \mathcal H_R$.\footnote{Mixed states generalising TFD were considered from a different perspective in \cite{Verlinde:2020upt}. A similar density matrix to the ones we consider was also recently considered for different purposes in \cite{Gao:2021tzr}.} For small departures from TFD, this is motivated by noting that the geometry obtained in \cite{Shenker:2013pqa} or \cite{Goel:2018ubv} doesn't depend on the details of the operators considered, so the bulk geometry can be related to a density matrix obtained by averaging over the operators.  Density matrices are usually written as a sum over states in $\mathcal H_L \otimes \mathcal H_R$; we refer to this as the {\it state frame} of the density matrix. The density matrices corresponding to different bulk geometries should have disjoint support in the space of quantum states. Different states drawn from the same density matrix have the same gravitational profile, but differ in microscopic data. 

We will argue that in considering two-sided correlations like $\langle \mathcal O_L \mathcal O_R \rangle$, it is also useful to rewrite the same density matrix as a sum of products of pairs of operators, one acting on $\mathcal H_L$ and the other on $\mathcal H_R$, We will refer to this way of writing as the {\it operator frame} of the density matrix. We can then view the density matrix as an object which takes an operator on $\mathcal H_R$ and gives us a corresponding operator on $\mathcal H_L$; that is we think of the density matrix in terms of operator conversion from one side to the other. If this operator conversion is close to diagonal, then two-sided correlations like $\langle \mathcal O_L \mathcal O_R \rangle$ will be of the same order as single sided correlators (like $\langle \mathcal{O}_R\mathcal{O}_R \rangle$). Thus, we argue that the density matrices of interest are those which 1) average in an appropriate way over microscopic details, 2) have disjoint support in the space of quantum states, and 3) have an approximately diagonal operator conversion for simple operators for which we expect to have gravitational duals. Going to the density matrix is essential in order to be able to formulate the last criteria, which goes beyond just entanglement. These general ideas are developed in section \ref{sec:gend}.

We will explore these questions explicitly in the context of the SYK model \cite{Sachdev:1992fk,Kitaev_talk,Maldacena:2016hyu}, and in a related model where the Majoranas are replaced by Pauli matrices (see for example \cite{Erdos:2014zgc}). The SYK model is a quantum mechanical system with $N$ Majorana fermions, with an all-to-all interaction of $p$ fermions at a time, where the strength of each such $p$-interaction is an independent random Gaussian. In the Pauli matrices model there are $N$ spin $1/2$ Hilbert spaces with a general random $k$-local interaction made out of Pauli matrices. In the large $N$ limit, the SYK model has a nearly-conformal symmetry in the IR, and fluctuations around it are described by a Schwarzian effective action, which is the same dynamics that describes JT gravity on $\text{NAdS}_2$ \cite{Maldacena:2016upp,Jensen:2016pah}. This provides a useful context for studying holographic relations where we can do some calculations directly in the quantum mechanical system. One of our main aims is to study semiclassical generalizations of the TFD from the boundary perspective, in contrast with earlier works which have focused primarily on calculation in the bulk holographic picture. We will mainly do calculations in a double scaled limit of SYK \cite{Erdos:2014zgc,Cotler:2016fpe}, where we take $p$ and $N$ to infinity with $p^2/N$ fixed, using the chord diagram approach to calculations in this limit discussed in \cite{Berkooz:2018qkz,Berkooz:2018jqr}, but the main conclusions carry over to the usual SYK large $N$ limit. The models and this calculational approach are reviewed in section \ref{sec:syk}. 

Part of going from a state on the doubled Hilbert space to desntity matrices on it, has to do with averaging over microscopic information (keeping macroscopic data of the wormhole, such as its length, fixed). An important clue for how to do so is the following. After averaging over couplings, the SYK model has an $O(N)$ symmetry under rotation of the basis of fermions, $\psi_i = O_i^j \psi_j$. When we consider two copies of the SYK model, because we have a single average over the couplings, the averaged system is invariant under a diagonal $O(N)$ symmetry which rotates both the left and right fermions simultaneously. There is a similar symmetry in the random spin model. We will argue that this rotation is an example of the kind of microscopic details we should average over in constructing duals of bulk wormholes - two configurations that differ by an $O(N)$ rotation will look the same gravitationally. Thus we will consider $O(N)$-invariant density matrices $\rho$, or more precisely, density matrices where the only terms which break the $O(N)$ symmetry are explicit insertions of the Hamiltonian.
In section \ref{sec:rho} we set up the specific class of density matrices we consider, and in section \ref{sec:CrlTwoSdd} we use the chord diagram approach to calculate two-sided correlation functions in this class of density matrices. We show that in a suitable limit the results obtained from our microscopic calculation reproduce the holographic calculation of correlations in a shockwave perturbation of TFD in \cite{Shenker:2013pqa}. 

In the following section \ref{sec:gend} we summarize the main points of the paper in more detail.

\section{Summary of main points - from TFD to density matrices on the doubled Hilbert space} 
\label{sec:gend}

We wish to consider generalisations of the thermofield double state from a microscopic perspective, and explore which ones could be dual to a semiclassical wormhole geometry in AdS spaces \cite{Maldacena:2004rf}. Our main aim is to study this explicitly in the context of the SYK model, where calculations in the microscopic quantum theory are (somewhat) tractable. But before setting up the details of our study in the SYK context, in this section we will set out the kind of generalisation we want to consider in a more general context and make contact with related work. 

We will consider states and density matrices in a bipartite quantum system with Hilbert space $\mathcal H_L \otimes \mathcal H_R$. We consider theories with decoupled dynamics, so the Hamiltonian is $H = H_L + H_R$; the connection between the two copies of the system comes only from entanglement in the quantum state.\footnote{This is in contrast to the work of \cite{Maldacena:2018lmt,Maldacena:2017axo}, which considered a coupling between the two theories such that a state close to TFD is the ground state, or \cite{Saad:2018bqo,Saad:2019lba} where two space connections were examined by looking at connected contribution to the spectral form factor.}

Our main points are that 1) for the purposes of understanding the gravitational dynamics of the wormhole it is useful to go to a density matrix on ${\cal H}_L\otimes {\cal H}_R$, and 2) there is a simple characterization of whether a density matrix corresponds to a semi-classical wormhole, which is not strictly the entanglement between the Right and Left spaces. {\it From now on when we say density matrix we will refer to a density matrix on ${\cal H}_L\otimes {\cal H}_R$}.

\subsection{A short review of deformations of the TFD}\label{sec:RevDTFD}

In \cite{Shenker:2013pqa}, small modifications of the TFD state were studied, introducing a perturbation by acting on one side with an operator $W(t) = e^{-iHt_w} \mathcal O e^{iHt_w}$, where $\mathcal O$ is some local operator. This ``timefold'' operator corresponds to evolving the TFD state at $t=0$ back into the past for a time $t_w$, inserting $\mathcal O$ and then evolving forward in time by $t_w$ back to $t=0$ to produce a modified state. In the limit of large $t_w$, the dual of this construction is a geometry with a lightlike shock propagating from the boundary at early times into the black hole. The back-reaction of this shock lengthens the wormhole. This investigation was extended to consider multiple shocks in \cite{Shenker:2013yza}. 

\begin{figure}
    \centering
    \includegraphics[page=1,width=0.45\linewidth]{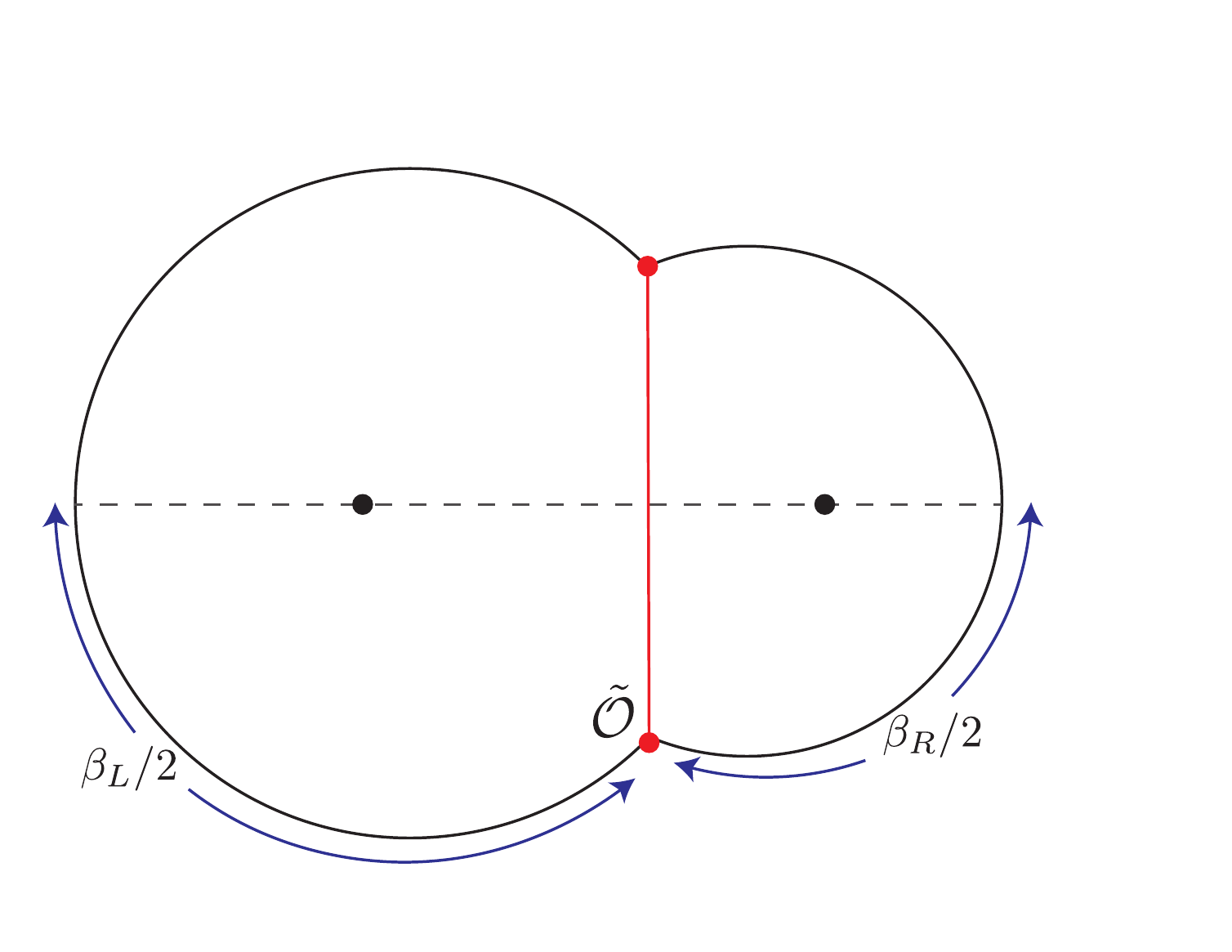}
    \includegraphics[page=2,width=0.45\linewidth]{Figures/BulkReconstruction.pdf}
    \caption{On the left, the construction of the PETS by a Euclidean path integral, and the dual bulk geometry. The $t=0$ slice of the Euclidean geometry provides initial data for the Lorentzian geometry on the right.}
    \label{fig:BulkRecon}
\end{figure}

A related investigation in \cite{Goel:2018ubv} considered what they termed the ``partially entangled thermal state'' (PETS)
\begin{equation} \label{pets}
|\bar{\mathcal{O}} \rangle = \sum_{m,n} e^{-\frac{1}{2} \beta_L E_m - \frac{1}{2} \beta_R E_n} \bar{\mathcal O}_{n,m} \Theta |m\rangle_L  |n \rangle_R,     
\end{equation}
where $\bar{\mathcal O}_{n,m}$ are the matrix elements of some local operator $\bar{\mathcal O}$ between energy eigenstates and $\Theta$ is an anti-unitary operator, for example CPT. The construction of this state is essentially a Euclidean variant of the previous construction: it's given by a Euclidean path integral with evolution over a period $\beta_R/2$ in Euclidean time, followed by insertion of the operator $\bar {\mathcal O}$, and further evolution over a period $\beta_L/2$ in Euclidean time, as depicted in figure \ref{fig:BulkRecon}. This state was introduced as an interesting generalization of both the TFD state and the ``thermal pure states'' considered in \cite{Kourkoulou:2017zaj}.

In \cite{Goel:2018ubv}, these states were studied from the spacetime dual perspective using JT gravity: it was argued that the dual of such states is a bulk geometry with a particle emitted into the bulk by the insertion of the operator $\bar{\mathcal O}$. They considered the regime where this particle is heavy (the dimension of the operator $\bar{\mathcal O}$ scales with the central charge in the holographic large central charge limit) so the emission of this particle back-reacts on the trajectory of the ``boundary particle'' in JT gravity, modifying the geometry. The modified geometry looks like a black hole from the perspective of each asymptotic region, but with a separation between the two horizons, producing a lengthened wormhole, see figure \ref{fig:BulkRecon}. In \cite{Goel:2018ubv}, the entanglement entropy and entanglement wedges of this bulk geometry were investigated using the replica trick, showing that the entanglement entropy is given by the smaller of the two horizon entropies. 

 A similar lengthening of the wormhole was also obtained in
 \cite{Balasubramanian:2014gla} when the state in the doubled theory was slightly rotated away from thermofield double, and in  
 \cite{Balasubramanian:2020coy} when entangling a non-gravitating system and a gravitating (with JT gravity) one, and including the backreaction on the geometry. 

\subsection{A convenient purification }

Usually, when going from a single to the two-sided discussion of a black hole, one purifies the thermal density matrix with another copy of the same Hilbert space. For us it will be more convenient to use a slightly different purification. Starting from the thermal density matrix on a single space $\rho(\beta)=\sum e^{-\beta E_i} |i\rangle \langle i|$, we purify it using an additional copy of the Hilbert space and a Hamiltonian with the same spectrum as the original Hamiltonian. This is true for another copy of ${\cal H}$ but it is equally true if we take ${\cal H}^\dagger$ for the second copy, and we will take our purifying space to be the latter.

That is, we just purify the density matrix by a state on 
${\cal H}_L^\dagger\otimes {\cal H}_R $  which we write as
\begin{equation} \label{eq:H2DaggerExample}
    |\psi(\beta)\rangle = \sum e^{-\beta E_i/2} |i\rangle_R \otimes {}_L\langle i | \in   {\cal H}_L^\dagger \otimes {\cal H}_R.
\end{equation}
This choice of purification arises naturally from a Euclidean path integral perspective. 
This is true in any dimension, but in the context of 2D gravity, this is particularly simple. Two sided black holes can be created by cutting a Euclidean space, whose boundary in the past is an open 1D surface (as in figure \ref{fig:BulkRecon}  on the left). This 1D boundary is can be chosen to have an orientation (any of the two will do). It is then natural to associate different orientations to the edges of this 1D surface, which translates into doubling ${\cal H}$ on one side by an ${\cal H}^\dagger$ on the other. One example of the usefulness of this choice is that we can modify a wormhole by the insertion of operators on this surface, and in this purification prescription the concatenation of operators action is straightforward and does not require any anti-unitaries. We can then generate a host of other states, such as states discussed in Fig. 15 of \cite{Goel:2018ubv}:
\begin{align}\label{VrlndB}
    e^{-\beta_3 H}\mathcal{O}e^{-\beta_2 H}\mathcal{O}e^{-\beta_1 H}
    =
    \raisebox{-.4\height}{\includegraphics[width=0.3\columnwidth]{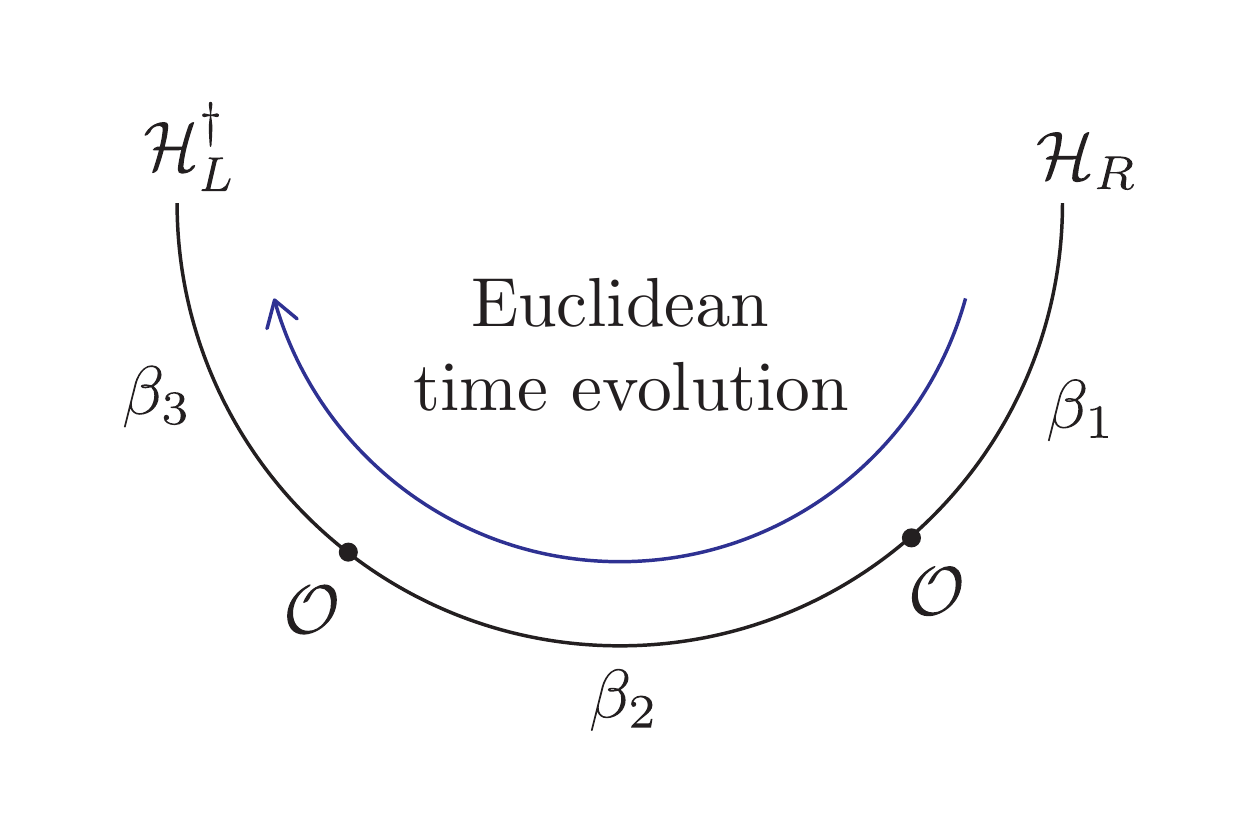}
    }
\end{align}

There is a one-to-one correspondence between this description using ${\cal H}_L ^\dagger\otimes {\cal H}_R $ and the description where we purify using ${\cal H}_L\otimes {\cal H}_R$ (adopted for example in \eqref{pets}) by mapping from $\mathcal{H}^\dagger$ to $\mathcal H$ as
\begin{align}
   \bra{n}_L \longleftrightarrow 
    \Theta \ket{n}_L,
\end{align}
where $\Theta$ is an anti-unitary isometry, for example  $CPT$. The need for this anti-unitary is because the time-evolution on ${\cal H}_L$ and ${\cal H}_L^{\dagger}$ are reversed in relation to each other. We will work in the ${\cal H}_L^\dagger \otimes {\cal H}_R$ purification scheme.

\subsection{Going to density matrices on the doubled Hilbert space}
\label{introdens}

The examples in section \ref{sec:RevDTFD} share the feature that they  focus on pure states on the doubled Hilbert space, which we take to be ${\cal H}^\dagger_L\otimes {\cal H}_R$. The goal of this paper is to show that certain questions clarify significantly when going further to {\it density matrices} on ${\cal H}^\dagger_L\otimes {\cal H}_R$. The idea is that deforming away from the thermofield double involves introducing an excitation in the bulk spacetime, but the excited wormhole geometry does not depend on many of the details of how we do it; that is, the same gravitational profile can be generated in various ways, and we can hence identify this geometry with some density matrix on ${\cal H}_L^\dagger\otimes {\cal H}_R$. 

A concrete example would be taking the state (\ref{pets}) of \cite{Goel:2018ubv} and averaging over the operator. The bulk geometry in \cite{Goel:2018ubv} depends only on the mass of the particle, that is on the operator dimension, and not on the details of which particular operator we insert. It therefore seems natural to consider a duality not between the bulk spacetime and an individual pure state of the form \eqref{pets}, but between the spacetime and a density matrix $\rho = \sum_i c_i|\bar{\mathcal O}_i \rangle \langle \bar{\mathcal O}_i |$, where we average over all operators with conformal dimensions in some window, with some appropriate weights.

The reverse direction is more interesting: suppose that we have a way of associating a density matrix to each geometric profile of a wormhole. Given a density matrix $\rho$ in some Hilbert space, here ${\cal H}^\dagger_R\otimes {\cal H}_L$, it defines a probability on the states in that Hilbert space\footnote{One way is to diagonalize the density matrix and say that we get an eigenvector with the associated probability. Another way is to use the density matrix to define a measure on the unit sphere in the Hilbert space.}.
We can then consider generic states by this measure. I.e., there would be states whose probability will decrease like $1/e^{S}$, where $S$ in the entropy of the density matrix, and there would be states where the probability will be much smaller. The former are the generic states in the density matrix. The dual of any of these generic states will look like the same gravitational wormhole.

Working with density matrices in the doubled Hilbert space will simplify the discussion, as it removes the dependence on some of the microscopic details. More interestingly it highlights a criterion for the existence of a semiclassical wormhole which is complementary to  entanglement, which we discuss next. 

\subsection{The state frame and the operator frame}

We will argue next that we can give a criteria (beyond entanglement) for when the wormhole is semi-classical in terms of this density matrix. The criteria reflects the more detailed structure of the algebra of operators in the theory. 

There are two complementary ways of viewing the density matrix, corresponding to different perspectives on the emergence of a wormhole in the bulk. A density matrix on ${\cal H}_L^\dagger\otimes {\cal H}_R$ is an object 
\begin{equation}\label{eq:DenMat2}
    \rho\in {\cal H}_L^\dagger\otimes {\cal H}_R\otimes {\cal H}_L\otimes {\cal H}_R^\dagger.
\end{equation}
We can write the density matrix as a sum over pure states in $\mathcal H^\dagger_L \otimes \mathcal H_R$. We will call this {\it the state frame}
\begin{equation}
    \rho = \sum_{\alpha\beta} M_{\alpha\beta} |\alpha\rangle \langle \beta |, \ \ \ \ |\alpha\rangle, |\beta\rangle \in {\cal H}_L^\dagger\otimes{\cal H}_R.
\end{equation}
As each state entagles left and right d.o.f, it emphasizes the entanglement structure between left and right. 

Alternatively we can write the density matrix as a sum of a product of operators acting on $\mathcal H^\dagger_L$ and on $\mathcal H_R$. We will call this {\it the operator frame}, and it emphasizes the correlators $\langle \mathcal O_L \mathcal O_R \rangle$ between operators on the two boundaries. The relation between the sum of states and the correlator is well understood for the TFD, but in the generic case, we argue that the existence of a semiclassical wormhole is better addressed from the latter perspective.  Just requiring entanglement (the first point of view) does not imply large correlation functions. 

In more detail, the {\it operator frame} is obtained instead by expanding \eqref{eq:DenMat2} as a sum of products of terms, one in ${\cal H}_R\otimes {\cal H}_R^\dagger$ and one in ${\cal H}_L^\dagger\otimes {\cal H}_L$ - i.e. a sum of products of operators, one acting on ${\cal H}_R$ and one on ${\cal H}_L^\dagger$:
\begin{equation}\label{eq:DnstOpsF}
    \rho= \sum_{a,b} C_{ab} {\cal O}_R^a {\cal {\tilde O}}^b_L ,
\end{equation}
where ${\cal O}^a_R$ are a list of operators acting on ${\cal H}_R$, and ${\tilde O}^b_L$ the similar list of operators acting on ${\cal H}^\dagger_L$. Given a rich enough set of operators we can expand any $\rho$ like this\footnote{Actually, for gravity purposes we care about the part of the expansion where these are single trace operators or otherwise objects that go through the wormhole without perturbing it much.} with some coefficients $C_{ab}$.
In this form, the density matrix implies an {\it operator conversion} between the two spaces, to which we turn next. 

\subsection{Operator conversion/pairing}
\label{sec:conv} 

A common probe of the connectedness of the bulk spacetime geometry is to consider the two-point function of a simple operator $\langle \mathcal O_L \mathcal O_R \rangle$; if the dimension of the operator $\mathcal O$ is much larger than one (but not scaling with the central charge, so we can neglect its back-reaction), the boundary correlator can be calculated in the bulk using the geodesic approximation, and the correlator will measure the (regularised) length of a geodesic connecting the two boundaries. Thus, such two-point functions being order one\footnote{Order one meaning that it does not scale with central charge in the large central charge limit, for conventionally normalized operators where the two-point function $\langle \mathcal O_L \mathcal O_L \rangle$ is fixed in the large central charge limit.}, and being independent (at leading order) of the details of the operator considered, is a useful diagnostic of the existence of a bulk dual. 

Consider now \eqref{eq:DnstOpsF}. This object gives a pairing of an operator from the right and from the left. 
 If we compute a two point function then
\begin{equation}
    \langle \mathcal O_R \mathcal O_L \rangle = \sum_{a,b} C_{ab} 
    \langle \mathcal O_R {\cal O}_R^a \rangle_R    
    \langle {\cal {\tilde O}}^b_L \mathcal O_L \rangle_L.
\end{equation}
Concrete examples of this are discussed in section \ref{sec:rho}.
A semiclassical, weakly coupled, wormhole corresponds to a density matrix which has $C_{ab}\propto \delta_{ab}$, as will be explained later.

Another way of phrasing it is as a {\it conversion} of the operators in ${\cal H}_R$ into operators on ${\cal H}_L$. A density matrix on the doubled Hilbert space literally plucks an operator from space R and inserts it into space L. To see this take \eqref{eq:DnstOpsF} and consider it as a map of operators
\begin{equation}
     \rho: \mathcal O_R \rightarrow \mathcal O_L= \sum_{a,b} C_{ab} 
    \langle  \mathcal O_R {\cal O}_R^a \rangle_R    
    {\cal {\tilde O}}^b_L.
\end{equation}
 The interpretation is that if we start with an operator in $R$ and move it through the wormhole, we will obtain the image $\mathcal O_L$ on the left hand space. For example if we start with the particle that corresponds to $\mathcal O_R$ and move it through the wormhole it will correspond to the superposition of particles encoded in the image $\mathcal O_L$.

Clearly here again we see that a weakly coupled, semiclassical wormhole, is characterized by $C_{ab}$ being close to diagonal (for any $\mathcal O_R$ that corresponds to a low energy field). The size of the diagonal $C_{ab}$ tells us what is the length of the wormhole. 

\subsection{Invariance}
\label{invsum}

We argued before that the key advantage of going to a density matrix is that it clumps together many pure states, so we can avoid considering features of the pure state which are not relevant to the gravitational profile of the wormhole. In cases like the SYK model this can be made more precise. The theory has an $O(N)$ action which rotates the fermions, and it is broken by the random couplings in the Hamiltonian. After we average over couplings, this symmetry is restored. If we consider two copies of the theory, since we consider a single average over couplings, it is only the diagonal $O(N)$ symmetry which acts on both copies which is restored.  Since gravity captures averaged quantities,  it is sensitive to objects that are invariant under $O(N)$, so it is natural to consider density matrices which are invariant under $O(N)$ - for example, we can average \eqref{pets} over all operators related by the $O(N)$ rotations. Note that this is only possible once we consider density matrices; the Hilbert space $\mathcal H_L^\dagger \otimes \mathcal H_R$ has just one singlet state which is invariant under the diagonal $O(N)$ symmetry (we will denote it by $|s\rangle$). But there are non-trivial $O(N)$ invariant density matrices, allowing us to focus on the gravitational data.

Technically, we will not restrict strictly to $O(N)$ invariant density matrices; we will also consider density matrices where the $O(N)$ symmetry is broken only by explicit insertions of factors of the Hamiltonian. For example, the infinite temperature limit of the thermofield double state is the $O(N)$ invariant singlet state $|s\rangle$, but in $|TFD \rangle \langle TFD| = e^{-\beta H/2} |s \rangle \langle s| e^{-\beta H/2}$, the $O(N)$ symmetry is broken by the insertions of the Hamiltonian. We are interested in objects which generalise the thermofield double state so we allow breaking of $O(N)$ by the Hamiltonian in our more general density matrices as well. 

This can probably be generalized to higher dimensions, with the $O(N)$ symmetry restrictions replaced by algebraic structures in the theory such as the OPE, but we leave this for future work.

\section{The models, symmetry and chord diagrams}
\label{sec:syk}

In this section we review the two models that we use - the SYK Majorana fermion \cite{Kitaev_talk,Maldacena:2016hyu,Sachdev:1992fk} and the Pauli matrices model \cite{Erdos:2014zgc,Berkooz:2018jqr,Berkooz:2018qkz}. We will be interested in the case of the two sided black hole and hence the space is doubled. We will then review the chord diagram techniques for these models, in the double scaling limit. Our main conceptual points do not rely on the double scaling limit, but it will be useful computationally.

\subsection{Two decoupled SYK models and their symmetry} \label{sec:MajoranaModel}

The Sachdev-Ye-Kitaev model is a quantum mechanical system in 0+1 dimensions, constructed of $N$ Majorana fermions with all-to-all interactions, and random (disordered) couplings. Denote the Majorana fermions by $\chi^{i}$, $i=1,\cdots,N$. These satisfy the algebra 
\begin{align}
    \{\chi^i,\chi^j\}=2\delta_{ij}.
\end{align}
For some integer $p\in 2\mathbb{N}$, we define the ordered multi-index $I\equiv(i_1,\cdots,i_p)$ (with $1\leq i_1<i_2<\cdots<i_p\leq N$), and the Majorana string $\chi^I=\chi^{i_1}\cdots\chi^{i_p}$. 
The Hamiltonian of the system is given by
\begin{align} \label{eq:Majorana_SYK}
    H = i^{p/2}{\binom{N}{p}}^{-1/2}\mathcal{J}\sum_{|I|=p}J_I\chi^I,
\end{align}
where $\mathcal{J}$ is the coupling constant, but by choosing the appropriate scale we set this to one.
The sum is over all possible multi-indices of length $p$. The $J$'s are taken to be Gaussian random variables with zero mean and normalized variance. We denote the expectation value over these by $\left<\cdot\right>_J$, such that we have
\begin{align} \label{eq:Gaussian_Coeffs}
    \left<J_{I}J_{I'}\right>_J=\delta_{I,I'}.
\end{align}
This is a slightly different normalization than, say, in \cite{Maldacena:2016hyu} but going between the conventions is straightforward.

Since we would like to study the dual of the wormhole geometry, which has two boundaries, we are going to deal with two decoupled copies of this model, correlated by the fact that we have the same random coupling on both sides. We label the two boundaries as $L$ and $R$ (left and right), and mark it as a subscript.
The Hamiltonians are the same on the two spaces, i.e., 
\begin{align} 
\begin{cases}
    H_L = i^{p/2}{\binom{N}{p}}^{-1/2}\sum_{|I|=p}J_I\chi^I_L,\\
    H_R = i^{p/2}{\binom{N}{p}}^{-1/2}\sum_{|I|=p}J_I\chi^I_R \ ,
\end{cases}
\end{align}
and we take the Majorana fermions on the two sides to anticommute with each other, \it i.e., \rm 
$\left\{\chi_L,\chi_R\right\}=0$.

% but we can also redefine the fields to take them anti-commuting, as
% \begin{align}
%   {\tilde \chi}_L=(-1)^{F_R}\chi_L,\ \  
%     \left\{{\tilde \chi_L} ,(-)^{F_L}\chi_R\right\}=0.
% \end{align}
% We will use the anti-commuting fermion convention. {\bf MB DO WE NEED ALSO THE $(-1)^{F_L}$?} \nb{We never really used it, but this is aligned with Maldacena's convention in Maldacena-Kurokulu, so it's probably better to write it here rather than pick sides in the fight}

We will be intersted in states in the doubled theory. There is an independent $O(N)$ symmetry rotating each set of Majorana fermions, which is broken by the Hamiltonian. The average over the random couplings restores the diagonal $O(N)$ symmetry which rotates both $\chi^i_L$ and $\chi_R^i$ at the same time (it is only the diagonal symmetry which is restored as the couplings in the two theories are correlated).

\subsection{Random spin model} \label{sec:RandomSpin}

The random spin model in 0+1 dimensions is a close relative of the SYK model, where spin $1/2$ degrees of freedom assume the role of the Majorana fermions. The relevant operators that appear in the Hamiltonian are then Pauli matrices acting on the various spins, i.e., $\sigma_{i}^{a_i}$, $i=1,\cdots,N$, and $a_i=1,2,3$. Let $e=(i_1,\cdots,i_p)$ be a vector of length $p$ of distinct integers defining a subset of the $N$ sites (again we take $1\leq i_1<\cdots<i_p\leq N$), and let $a=(a_1,\cdots,a_p)$ be a second vector of length $p$, with $a_i=1,2,3$. Denoting the pair $(e,a)$ by $I$, we define $\sigma_I=\sigma_{(e,a)}=\sigma_{i_1}^{a_1}\cdots\sigma_{i_p}^{a_p}$. The Hamiltonian is now
\begin{align} \label{eq:RandomSpinHamiltonian}
    H_{\text{random spin}} = 3^{-p/2}{\binom{N}{p}}^{-1/2}\mathcal{J}\sum_{|I|=p}J_I\sigma_I, 
\end{align}
where again the sum runs over all possible choices for $I$, and $\mathcal{J}$ sets an energy scale which we normalize to $1$. As in section \ref{sec:MajoranaModel}, the $J_I$'s are taken to be independent random Gaussian variables with zero mean and unit variance, such that we have (\ref{eq:Gaussian_Coeffs}).

In the IR at finite $p$, the model was argued on quite robust grounds to not support a non-fermi liquid phase \cite{Baldwin:2019dki}. 
In fact, a related model, at fixed $p$, is reviewed in \cite{Anous:2021eqj}, and exhibits a plethora of complicated IR behaviors. 
However, we will be interested in the double scaling limit $p\propto \sqrt{N}$ where it shows the same universal dynamics as the double scaled SYK model which has nearly conformal physics in the IR.

We take two decoupled random models \eqref{eq:RandomSpinHamiltonian} but with the same random coupling, as in the Majorana SYK case.
The Hamiltonian we are interested in is therefore\footnote{Actually, a refinement is needed for the left operator to act on ${\cal H}^\dagger$ - see appendix \ref{sec:indices}.}
\begin{align} 
\begin{cases}
    H_L = 3^{-p/2}{\binom{N}{p}}^{-1/2}\sum_{|I|=p}J_I\sigma^I_L,\\
    H_R = 3^{-p/2}{\binom{N}{p}}^{-1/2}\sum_{|I|=p}J_I\sigma^I_R \ .
\end{cases}
\end{align}
For later reference, we can clump together degrees of freedom associated with spin $i$ from the left and from the right Hilbert space into a single four dimensional Hilbert space, which we denote as $\mathcal{H}^i$.

The symmetry of the system after averaging over the random coupling is $SU(2)^N\rtimes S_N$, where each $SU(2)$ acts on $\mathcal{H}^i$ (which is made out of an singlet and a triplet under this action), and $S_N$ is the permutation of the $N$ species of Pauli matrices on the left and the right boundaries at the same time.

\subsection{Probe operators} \label{sec:OpDef}

Once we map the set of states that correspond to some wormhole, we would like to start probing it. A set of convenient probes are two sided correlation functions, but we still need to choose of which operators.    

The operators that we will choose as probes are random operators. A random operator $M$ of length $\tilde{p}$ is defined to be
\begin{align} \label{eq:OpDef}
    M = \begin{cases}
    3^{-\tilde{p}/2}{\binom{N}{\tilde{p}}}^{-1/2}\sum_{|I|=\tilde{p}}\tilde{J}_I\sigma_{I} & \text{Random spin model}
    \\
    i^{\tilde{p}/2}{\binom{N}{\tilde{p}}}^{-1/2}\sum_{|I|=\tilde{p}}\tilde{J}_I\Psi_{J} & \text{Majorana SYK},
    \end{cases}
\end{align}

where the couplings $\tilde{J}_I$ are again random Gaussians with zero mean and normalized variance, 
\begin{align}
    \left<\tilde{J}_I\tilde{J}_{I'}\right>_{\tilde{J}} = \delta_{I,I'},
\end{align}
where $\left<\cdot\right>_{\tilde{J}}$ denotes the ensemble average. Notice that the coefficients $\tilde{J}_I$ are independent of the coefficients $J_I$.

We could consider instead the operators to be ``monomials" $\sigma_I$ or $\Psi_J$ for a specific $I$ or $J$ - nothing would have changed in that case, since we can view the random operator above as computing the properties of a generic monomial. 

There is, however, a deeper reason for choosing such operators. Consider the case that we have some higher dimensional theory that flows to a (near) extremal background of the form AdS$_2 \times M$, and suppose that this IR can be thought of as an SYK model in some variables (which may not be related to the original UV d.o.f). The set of observables correspond to fields defined in the UV of the entire background. The local energy-momentum is one of them, but it is typically part of a whole tower of similar ("single trace") operators. If the Hamiltonian in the full theory flows to some random Hamiltonian on the near extremal BH degrees of freedom, then we can expect the same to be true of all other UV operators. So the natural probes are random operators in a statistical class similar to the Hamiltonian, except that that we allow them to have different quantum numbers, which in this case is just the conformal dimension/length of the operator ${\tilde p}$ (more precisely, the conformal dimension of the operator in the IR is ${\tilde p}/p$, where $p$ is the length of the Hamiltonian). 

\subsection{The double scaled limit and chord diagrams}

In this work, we will be interested in computing quantities of the form $Z=\left<\Tr(e^{-\beta H})\right>_J$, where the subscript J means averaged over the ensemble of random couplings. We will also be interested in the 2-sided correlation functions in the density matrices discussed above.

This computation will prove to be especially simple in the limit in which we take $N\to\infty$, and scale $p\sim\sqrt{N}$. We refer to this limit as the \textit{double scaled limit}.  Namely, we keep fixed the parameter
\begin{align} \label{eq:lambda}
    \lambda = \begin{cases} 
    \frac{2p^2}{N} & \text{Majorana SYK} \\
    \frac{4}{3}\frac{p^2}{N} & \text{Random spin model},
    \end{cases}
    \qquad q \equiv e^{-\lambda}.
\end{align}
We can define analog quantities to $\lambda,q$ for probe operators other than the Hamiltonian by
\begin{align}
\tilde{\lambda} = \begin{cases}
\frac{2p\tilde{p}}{N} & \text{Majorana SYK}
\\
\frac{4}{3}\frac{p\tilde{p}}{N} & \text{Random spin model,}
\end{cases}    
\end{align}
where ${\tilde p}$ is the length of the new operator, and for both models we define $\tilde{q},\tilde{\ell}$ by
\begin{align} \label{eq:q_tilde}
    \tilde{q} = e^{-\tilde{\lambda}}= q^{\tilde{\ell}}, \qquad {\tilde{\ell}}={\frac{\tilde p}{p}}.
\end{align}

The main advantage of working in this limit is that it enables us to use \textit{chord diagrams} in order to compute various quantities. These computations have already been performed in \cite{Berkooz:2018jqr,Berkooz:2018qkz}, and the following summary of the methods used will serve as the basis for the computation in the rest of the work.

\begin{figure}
    \centering
    \includegraphics[width=0.4\textwidth]{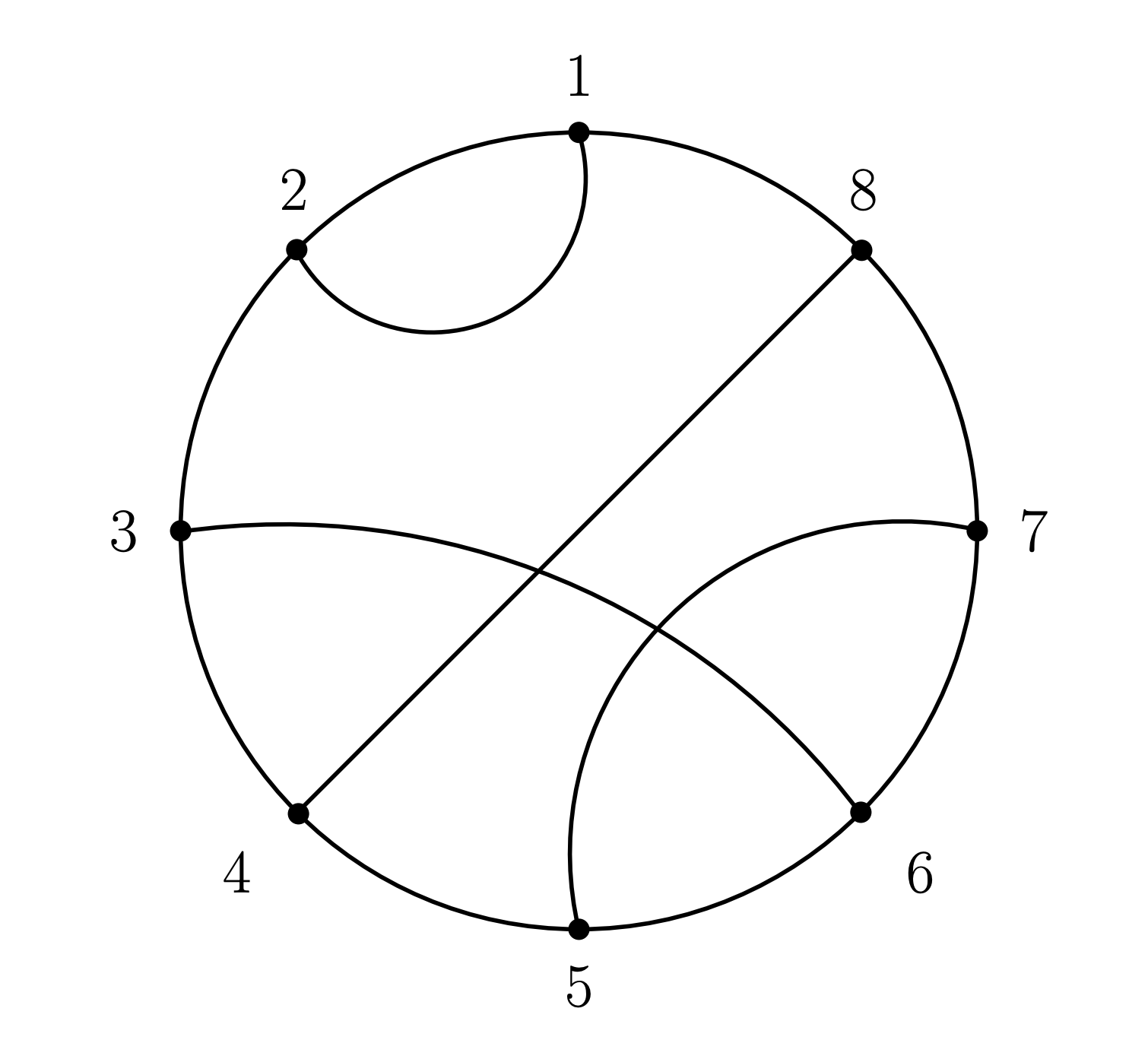}
    \caption{An example of a chord diagram for $k=8$. With the numbering in the figure, this diagram means that $I_1=I_2,I_3=I_6,I_4=I_8$ and $I_5=I_7$.}
    \label{fig:SingleSided_CD_Example}
\end{figure}

As an example, we'd like to compute the partition function\footnote{We will present for the formulas for correlations functions when we need them.}. First we Taylor expand $Z=\sum_{k}\frac{(-\beta)^k}{k!}\left<\Tr(H^k)\right>_J$, and then we evaluate each of the moments
\begin{align} \label{eq:Partition_Moment}
    m_k \equiv \left<\Tr(H^k)\right>_J. 
\end{align}
Since the coefficients $J_I$ are random Gaussians, upon taking the ensemble average we Wick contract them. Together with the cyclic structure of the trace, this allows us to represent the moment $m_k$ as a sum of \textit{chord diagrams}, as in figure \ref{fig:SingleSided_CD_Example}. The chord diagram then determines the arrangement of repeating Majorana fermions or Pauli matrices inside the trace. A little combinatorics is then needed to show that in the double scaled limit, the value of every such chord diagram is given by the number of chord intersections, so 
\begin{align} \label{eq:TChordPF}
    m_k = \text{dim}({\cal H})\times \sum_{\text{CD}(k)}q^{\# \text{intersections}},
\end{align}
where $\text{CD}(k)$ represents chord diagrams with $k$ nodes, and \#intersections is the number of pairwise intersection of chords in the diagram. This sum can be evaluated using a transfer matrix method in which one goes linearly along the circle, adding a new chord or closing a chord (with appropriate weight) at each node. A full explanation of the chord techniques is given in appendix \ref{sec:DSLimitAndCD}.

The advantage of this limit is that, in our application, it ``combinatorizes" the geometric relation between the two universes. In the one side case, generally, the chords are the objects that carry correlations of the different terms in the Hamiltonian from one time to another, i.e, from one boundary point to another in the dual GR picture. When we have two spaces we can have chords stretching between pieces of the Hamiltonian on the two different sides. This is a combinatorial manifestation of a bridge in spacetime that forms between the two systems. 

\section{Microscopics of wormhole density matrices}
\label{sec:rho}

In this section we discuss the density matrices that are relevant to the Pauli matrices model (sections \ref{sec:SpinSetUp} to \ref{sec:TrnsMtrc}) and to the Majorana case (section \ref{sec:MajSetUp}). We discuss their form in the state frame and in the operator frame, and their entanglement, which is easily read form one frame, vs. their length, which is easily read from the other frame.

\subsection{Density matrices in the random spin model}\label{sec:SpinSetUp}

Let us set up the states and density matrices we want to consider in the random spin model. We will begin by setting up our conventions with a single spin, and then go on to consider $N$ spins and the limit of large $N$. 

\subsubsection{Hilbert space structure of the two qubit system}
\label{sec:SnglSpn}

\paragraph{Basis of pure states:}

As a toy model, let us study a double copy of a system with a single spin. We think of the two copies as living on the left and the right boundaries.
Remembering that our random spin model had the symmetry $SU(2)^N\rtimes S_N$ after ensemble averaging, we will construct $SU(2)$ invariant density matrices.

 We will start with pure states. The doubled Hilbert space of a single spin has four states, which split into a singlet and a triplet under the diagonal $SU(2)$. Pure states on ${\mathcal{H}^\dagger_L}\otimes \mathcal{H}_R $ can be represented as operators from $\mathcal{H}_L$ to $\mathcal{H}_R$ (or operators on ${\cal H}$ since ${\cal H}_{L,R}$ are isomorphic). In the present case, the identification of states with operators is
\begin{align} %\label{eq:StateOp}
    \begin{cases} 
        \ket{s}\equiv \frac{1}{\sqrt{2}}\mathbbm{1}, & \\
        \ket{t,m}\equiv \frac{1}{\sqrt{2}}\tau^a, & \\
    \end{cases}
    \label{eq:defsinglet}
\end{align}
where $m$ takes values in $-1$, $0$, or $1$, corresponding to the three Pauli matrices $\tau^a$.
An element $g$ of the diagonal $SU(2)$ acts on the operators as $\mathcal{O}\mapsto g\mathcal{O}g^\dagger$,
so $\mathbbm{1}$ is a singlet and $\tau^a$ are a triplet under the action of $SU(2)$.

\paragraph{Invariant density matrix}

We are interested in the density matrix which is invariant under the diagonal $SU(2)$. The most general invariant density matrix is 
\begin{equation} \label{eq:PhysicalDM}
    G(\hat{A}) = (1-{\hat A}) |s\rangle \langle s| + ({\hat A}/3)  \sum_{m=0,\pm 1} |t,m\rangle \langle t,m|
\end{equation}
and positivity of the density matrix implies $0\leq {\hat A}\leq 1$. 

This object lives in $\bigl( {\cal H}_L\otimes {\cal H}_R^\dagger\bigr)\otimes\bigl({\cal H}_L^\dagger\otimes {\cal H}_R\bigr)$. Alternatively we can write in the crossed channel $(\mathcal{H}_L\otimes\mathcal{H}_L^{\dagger})\otimes(\mathcal{H}_R\otimes\mathcal{H}_R^{\dagger})$.
Let $(\1_{\sigma_{L,R}},{\vec\sigma}_{L,R})$ and $(\1_{\tau_{I,O}},{\vec\tau}_{I,O})$ be Pauli matrices and the identity, defined in the following spaces
\begin{align}\label{eq:PauliChnls}
\begin{split}
    & {\1_{\tau_I},\vec\tau_I}\in\mathcal{H}_L^\dagger\otimes\mathcal{H}_R,\qquad {\1_{\tau_O},\vec\tau_o}\in\mathcal{H}_R^\dagger\otimes\mathcal{H}_L,
    \\
    & {\1_{\sigma_R},\vec\sigma_R}\in\mathcal{H}_R\otimes\mathcal{H}_R^{\dagger}, \qquad {\1_{\sigma_L},\vec\sigma_L}\in\mathcal{H}_L\otimes\mathcal{H}_L^{\dagger}.
\end{split}
 \end{align}

Using this notation, we can write the correspondence
\begin{align}
    \frac{1}{2}\1_{\tau_I}\1_{\tau_O} \leftrightarrow \ket{s}\bra{s},\qquad \frac{1}{2}\tau_I^m\tau_O^m\leftrightarrow \ket{t,m}\bra{t,m},
\end{align} 
so the density matrix becomes
\begin{align} \label{GAhat}
        G(\hat{A}) = \frac{1-{\hat A}}{2}\1_{\tau_I}\1_{\tau_O}+\frac{\hat{A}}{6}\vec{\tau}_I\cdot\vec{\tau}_O.
\end{align}
This is the form of the density matrix in the \emph{state frame}.

\paragraph{The {\it \bf operator frame}}
We can translate this to the crossed channel using the Fierz identity
\begin{align}
        \1_{\sigma_L}\1_{\sigma_R}+\vec{\sigma}_L\cdot\vec{\sigma}_R = 2\1_{\tau_I}\1_{\tau_o},\qquad \1_{\tau_I}\1_{\tau_o}+\vec{\tau}_I\cdot\vec{\tau}_O=2\1_{\sigma_L}\1_{\sigma_R}.
\end{align}
This gives us the density matrix in the crossed channel notation
\begin{align} \label{eq:CrossedChDM}
    G(\hat{A}) = \frac{1}{4}(\1_{\sigma_R}\1_{\sigma_L}+A\vec{\sigma_R}\cdot\vec{\sigma_L}), \qquad A=1-\frac{4}{3}\hat{A},\qquad -{\frac{1}{3}}\leq A \leq 1.
\end{align}
This is the form of the density matrix in the {\it operator frame}.

In terms of the variable $A$, the density matrix has the singlet with probability $\frac{1+3A}{4}$, while each triplet has a probability of $\frac{1-A}{4}$.

\subsubsection{Invariant density matrix of the large $N$ random spin model (single shot)}
 
Let us now move to our main interest, which is the $N$ spin system. We take $N$ copies of the spin-1/2 system, i.e., ${\cal H}=(\mathbb{C}^2)^N$. 
The symmetry of the two-sided system here is, as we have already explained, ${\cal G}=SU(2)^N\rtimes S_N$, where each $SU(2)$ is the diagonal symmetry on the two copies of $\mathcal{H}^i$, and $S_N$ is the permutation of fermion species. 

The simplest $\mathcal G$-invariant density matrix is attained by taking the product of $N$ copies of the single spin density matrix above, with the same value $A$ for each spin:
\begin{align}\label{SnglSht}
    G(A)={\frac{1}{4^N}} \prod_{i=1}^N \left(\1_{\sigma_{R,i}}\1_{\sigma_{L,i}}+ A {\vec\sigma}_{R,i}\cdot{\vec\sigma}_{L,i} \right).
\end{align}
Our investigations will focus on this case, which we will refer to as {\it single shot} density matrices.  In appendix \ref{ent}, we consider a more general class of $\mathcal G$-invariant density matrices, where we take different values of $A$ for different spins, and argue that \eqref{SnglSht} is preferred as the largest entropy density matrix for fixed wormhole length. Thus, this represents the broadest coarse-graining over the microstates, and is a natural candidate for a dual of a smooth semi-classical geometry, which does not distinguish between microstates. In section \ref{sec:TrnsMtrc}, we discuss the generalization to multi shot density matrices, which we leave for future work.

To make contact with the thermofield double, we need to include thermal factors on the left and on the right to suppress contributions from high energy states, considering the density matrix 
\begin{equation}\label{eq:RandomSpinDM}
   G(A,\beta_L,\beta_R) = Z^{-1} e^{-\beta_L H_L/4 - \beta_R H_R/4} G(A) e^{-\beta_L H_L/4 - \beta_R H_R/4}
\end{equation}
where $Z$ normalizes it to have trace 1. We will use $G$ when discussing these specific density matrices, and we will reserve $\rho$ for discussing density matrices in general.

For $A=1$ (that is, $\hat A =0$), $G(1) = |s\rangle \langle s|$,  where $|s \rangle = \prod_{i=1}^N |s_i\rangle$ is the product of invariant singlet states, which is the purification of the infinite temperature thermal density matrix (i.e., the identity matrix on ${\cal H}_R$). Thus adding these thermal factors converts it into the thermofield double state with temperature $\beta=(\beta_L+\beta_R)/2$, as also discussed in section \ref{invsum}. For $A=1$ the density matrix depends only on $\beta_L+\beta_R$, but for a general $A$ it depends on both separately. 

Thus, \eqref{eq:RandomSpinDM} is an interesting direction to generalize the thermofield double. The $SU(2)^N\rtimes S_N$ symmetry is broken only by the explicit insertions of the Hamiltonian. As discussed in section \ref{introdens}, this density matrix can be thought of as taking the PETS state \eqref{pets} and averaging over a class of operators.
We will argue below that \eqref{eq:RandomSpinDM} are good candidates for the dual of semi-classical wormholes on the gravity side, where we keep the gravitational profile of the wormhole and forget about the microscopics of the particles that created it. 

\subsection{Properties of the single shot density matrices}\label{sec:PropSnglSht}

We now discuss several properties of the single shot density matrix: we comment on a matching property for $A=1$, calculate the entropy of these density matrices, briefly discuss the length of the corresponding bulk wormhole -- arguing that for $A<1$ the wormhole gets longer, as in previous studies of generalisations of the thermofield double -- and argue that the overlap of these density matrices for different values of $A$ vanishes in the large $N$ limit. 

\paragraph{TFD and matching property:}
For $G(1)$, we have
\begin{equation}\label{eq:ConversionPropertyHam}
    (H_L \otimes \1_R - \1_L \otimes H_R) G(1) =  G(1) (H_L\otimes \1_R - \1_L\otimes H_R)\ .
\end{equation}
This means that if we act with $H_R$ on $G$ then we can convert it into an action of $H_L$.  This can be viewed as the technical reason why the thermal state has zero length in the interior of the wormhole. In fact, this is more general. The length of the wormhole corresponding to $G(1)$ is zero for any operator, which means that something like \eqref{eq:ConversionPropertyHam} should hold for any operator and not just $H$. We will refer to this the {\it matching property}. It is easy to verify that for any Pauli matrix,
\begin{equation} \label{eq:ConversionPropertyPauli}
 \bigl( \sigma^{\mu}_{L}\otimes \1_{R} - \1_{L}\otimes \sigma^{\mu}_{R}\bigr)  G(1) = G(1) \bigl( \sigma^{\mu}_{L}\otimes \1_{R} - \1_{L}\otimes \sigma^{\mu}_{R}\bigr).
\end{equation}
This means that any operator on ${\cal H}_R$ made out of a string of Pauli matrices can be matched with (or fully converted into) an operator on the left Hilbert space. This is done with relative strength 1, which translates into a zero length wormhole. 

When we discuss the thermofield double with finite temperature, we conjugate $G(1)$ with $e^{-\beta_{L,R}H_{L,R}}$. The core of $G(1)$ still matches operators in left and right but now there is an additional contribution of the thermal suppression factors which we interpret as a contribution in Left and Right separately outside the horizon. 

This matching property is distinct from the general operator conversion we discussed in section \ref{sec:conv}, and is special for $G(1)$ and the thermofield double. When we consider  $G(A\not=1)$, we will not have such a matching between operators on the left and the right. We can always use the operator conversion property introduced in section \ref{sec:conv} to push an operator on the right through the density matrix to an operator on the left, but we can't in general find a (simple) operator acting on the density matrix on the left whose action matches that of an operator on the right as in \eqref{eq:ConversionPropertyHam}. 

Indeed, already for a general pure state in $\mathcal H^\dagger_L \otimes \mathcal H_R$ such a matching may not be possible; matching an operator on the left to an operator on the right requires sufficient entanglement, and will fail for example for product states. But when we consider density matrices, the matching problem is more severe: a relation like \eqref{eq:ConversionPropertyHam} is a four-index tensor equation. If we try to solve it for the left operator $\mathcal O_L$  given a right operator, we have far fewer variables than equations, and there is no general reason to expect a solution to exist. As we consider $A \neq 1$, we have a superposition of pure states in the density matrix; if the solution of matching for different states in the superposition is different, no solution will exist for the density matrix itself.

\paragraph{Entropy}
From \eqref{GAhat}, the single species density matrix $G(A)$ contains the singlet with probability $1-\hat{A}$ and each basis of the doublet $\hat{A}/3$, $\hat{A}=\frac{3}{4}(1-A)$. This means that the entropy of the large $N$ density matrix is
\begin{equation}
    S(A)= -N\biggl( {\frac{1-3A}{4}} \log{ {\frac{1-3A}{4}} } + 3{\frac{1+A}{4}}  \log {\frac{1+A}{4}} \biggr)
\end{equation}

If we take $A$ to be fixed in the large $N$ limit, the entropy scales like $N$. This is too large because summing over $e^{cN}$ states is expected to correspond to additional horizons embedded in the background. In the explicit calculations, we will study the models in a double scaling limit of large $N,p$, with fixed $p^2/N$. We will find (see next section) that the density matrices of interest in the double scaling limit have $A=1-a/p$ for fixed $a$, i.e, $A=1-{\cal O}(1/\sqrt{N})$, in which case the entropy is
\begin{equation}\label{eq:DnstMtrxEnt}
    S(A)= -{\frac{3Na}{4p}} \cdot\biggl( \log\bigl( {\frac{a}{4p}}\bigr)-1\biggr)
  \end{equation}
which scales like $\sqrt{N}\log(N)$.
The same argument holds qualitatively for the finite $p$ model (see section \ref{sec:LenghWH}).

The origin of this entropy is the following: Recall that in order to obtain a wormhole we are inserting some source which backreacts on the geometry. The entropy is just the entropy associated with this source. We will see below that $a$ also determines the length of the wormhole, so in this case there is a straightforward relation between the entropy of the states that can generate a specific geometry and its length. This might be related to how one identifies complexity in the bulk \cite{Brown:2015bva,Susskind:2014rva}, but for this one would have to make a more precise relation between complexity and the entropy in \eqref{eq:DnstMtrxEnt}, as well as study it for more general density matrices. 

When we turn on temperature, $G(A)$ gets dressed to become $G(A,\beta_L,\beta_R)$.
Since the dressing is by conjugating $G(A)$ the basic structure of the gluing remains the same. Actually, a formal number of singlets and triplets can be defined in the dressed operator, by computing the expectation value of an appropriate projection operator but we will not pursues this further in the current paper. We expect however that the entropy will pick up some $\beta_L,\beta_R$ dependence but that the $N$ scaling will remain the same. 

\paragraph{Length of the wormhole}
We will argue that the length of the throat is proportional to $-\log(A)$. A more precise discussion of the length of the wormhole will be given in  section \ref{sec:CrlTwoSdd}, but for now let us give the gist of the argument. 

On the gravity side, if we have an operator which corresponds to a particle of mass $m$, then a two sided correlator (at time zero on both sides) will receive a contribution of the form 
\begin{equation}\label{eq:TwoSddLngth}
    \langle M_L(t_L=0) M_R(t_R=0) \rangle \propto e^{-mL},
\end{equation} where $L$ is the length of the wormhole, and $M_L$($M_R$) is the insertion of the operator on the left (right). For simplicity we will take $m\gg1$ such that $mR_{AdS}\sim \Delta$, the dimension of the operator. 

We realize an operator with conformal dimension $\Delta$ as a random operator of length ${\tilde p}=\Delta \cdot p$, as discussed in \ref{sec:OpDef}.
We would like to identify a contribution of the form \eqref{eq:TwoSddLngth} to the two sided correlator when evaluated with $G(A),\ A\not=1$. I.e., in the QM side we compute
\begin{equation}
    \langle M_L(0) M_R(0)\rangle_{G(A)} = \Tr_{{\cal H}^\dagger_L\otimes {\cal H}_R}\bigl( G(A) M_L(0)M_R(0) \bigr).
\end{equation}
and focus on terms which look like the RHS of \eqref{eq:TwoSddLngth}.

We explain the computation a bit more in section \ref{sec:CrlTwoSdd}, but the structure is clear. To get a non-zero answer, we need to have the same monomial of Pauli matrices both on $M_L$ and $M_R$ - say that it is characterized by an index set ${\tilde I}$ (of length ${\tilde p}$). For each index $i_1,...i_{\tilde p}$ in the set ${\tilde I}$ we need to use the $A\sigma_L\sigma_R$ term in that index to get a non-zero answer. This means that we get a contribution of the form 
\begin{equation}
    \left<M_L(0)M_R(0)\right>_{G(A)}\propto A^{\tilde p}=e^{m R_{AdS} \cdot p \cdot \log(A)}\ .
\end{equation}
We equate this with $e^{-mL}$ to obtain that $L/R_{AdS}\propto -p\cdot \log(A)$. This is the case for any $p$. Thus, we see that taking $A<1$ decreases the value of the two-sided correlator, corresponding to a longer wormhole in the gravitational dual. In the double scaling limit we consider in the explicit calculations in the next section,  $A=1-a/p$ for fixed $a$, and hence $L/R_{AdS}\sim a$.

\paragraph{Overlaps of $G(A)$'s:}
Providing the gravitational data for any semi-classical gravitational background will determine a density matrix. But since different semiclassical wormholes correspond to different quantum states, then it should be that the overlap in the support of two such measures is small. One can verify that this is the case. Consider $G(A_1)$ and $G(A_2)$ for $A_1\not=A_2$. Since $A$ determines the length then these two density matrices corresponds to wormholes with different length - i.e, they correspond to different semiclassical geometries that are macroscopically different. Recall that for each species $G(A_i)=\alpha_i P_s + \beta_i P_t$, where $P_{s,t}$ are projection operators on the singlet/triplet ($\alpha_i+3\beta_i=1$), then a reasonable measure of the overlap of the density matrices\footnote{$G(A)^{1/2}\propto \prod (a_i^{\frac{1}{2}} P_{s,i} + b_i^{\frac{1}{2}}P_{t,i})$} is given by
\begin{equation}
    \Tr ( G(A_1)^{1/2} G(A_2)^{1/2} )= (\alpha_1^{1/2}\alpha_2^{1/2}+3\beta_1^{1/2}\beta_2^{1/2})^N \rightarrow 0 
\end{equation}
in the large $N$ limit for $(\alpha_1,\beta_1)\not= (\alpha_2,\beta_2)$.
Actually, in the double scaling limit we have to be a little more careful since $\alpha_i\sim 1-a_i/p$, but the overlap still goes to zero.

\subsection{Transition matrices and multi shot spaces}\label{sec:TrnsMtrc}

{\bf Transition Matrices:} Our starting point was to consider an object in $\rho \in (\mathcal H_L^\dagger \otimes \mathcal H_R ) \otimes (\mathcal H_L\otimes \mathcal H_R^\dagger) $ either as a pairing of operators, one on $\mathcal H_R$ and one on $\mathcal H_L^\dagger$ (the operator frame) or as a density matrix on $\mathcal H_L^\dagger\otimes \mathcal H_R $ (the state frame). The semiclassical nature of a wormhole manifests itself simply in the first approach, and the entanglement structure in the second approach.

In the state frame $\rho=\sum_\alpha c_\alpha |\psi_\alpha\rangle \langle \psi_\alpha |$ where $|\psi_\alpha\rangle$ are some states in $\mathcal H_L^\dagger \otimes \mathcal H_R $. The sum over $\alpha$ averages all the ways that we build a wormhole with a fixed given gravitational profile. It is clear, however, that we can just as well build a mixed state transition matrix (or Pseudo-entropy \cite{Nakata:2020luh})
\begin{equation}\label{eq:TrnsMtrx}
    \rho=\sum_\alpha c_\alpha |\phi_\alpha\rangle \langle \psi_\alpha |
\end{equation}
where $|\phi_\alpha\rangle$ are a distinct set of states than $|\psi_\alpha\rangle$. The only requirement that we impose is that \eqref{eq:TrnsMtrx} has no explicit violations of the symmetry other than by insertions of the Hamiltonian.  

The simplest example is the following. Consider \eqref{eq:RandomSpinDM} or \eqref{eq:MajRho}. For $\beta_1=\beta_2$ it describes, in Minkowski space, a symmetric $L\leftrightarrow R$ configuration with a massive particle sitting in the middle. This is a density matrix of the form $\sum_\alpha c_\alpha |\psi_\alpha\rangle \langle \psi_\alpha |$. Suppose we want to change its initial entry point and its final point - this would be of the form \eqref{eq:TrnsMtrx}. This would then be given by the expression 
\begin{align} \label{eq:Non_Symm_dnsty_mtrxA}
    G \equiv e^{-(\beta_1+4it_p) H_1/4-(\beta_2+4it_p) H_2/4}G(A)e^{-(\beta_1-4it_f) H_1/4-(\beta_2-4it_f) H_2/4}.
\end{align}
where $t_f,\ t_p$ parameterize the entry/deprature point of the massive particle. 

\begin{figure}
    \centering
    \includegraphics[width=0.7\textwidth,page=4]{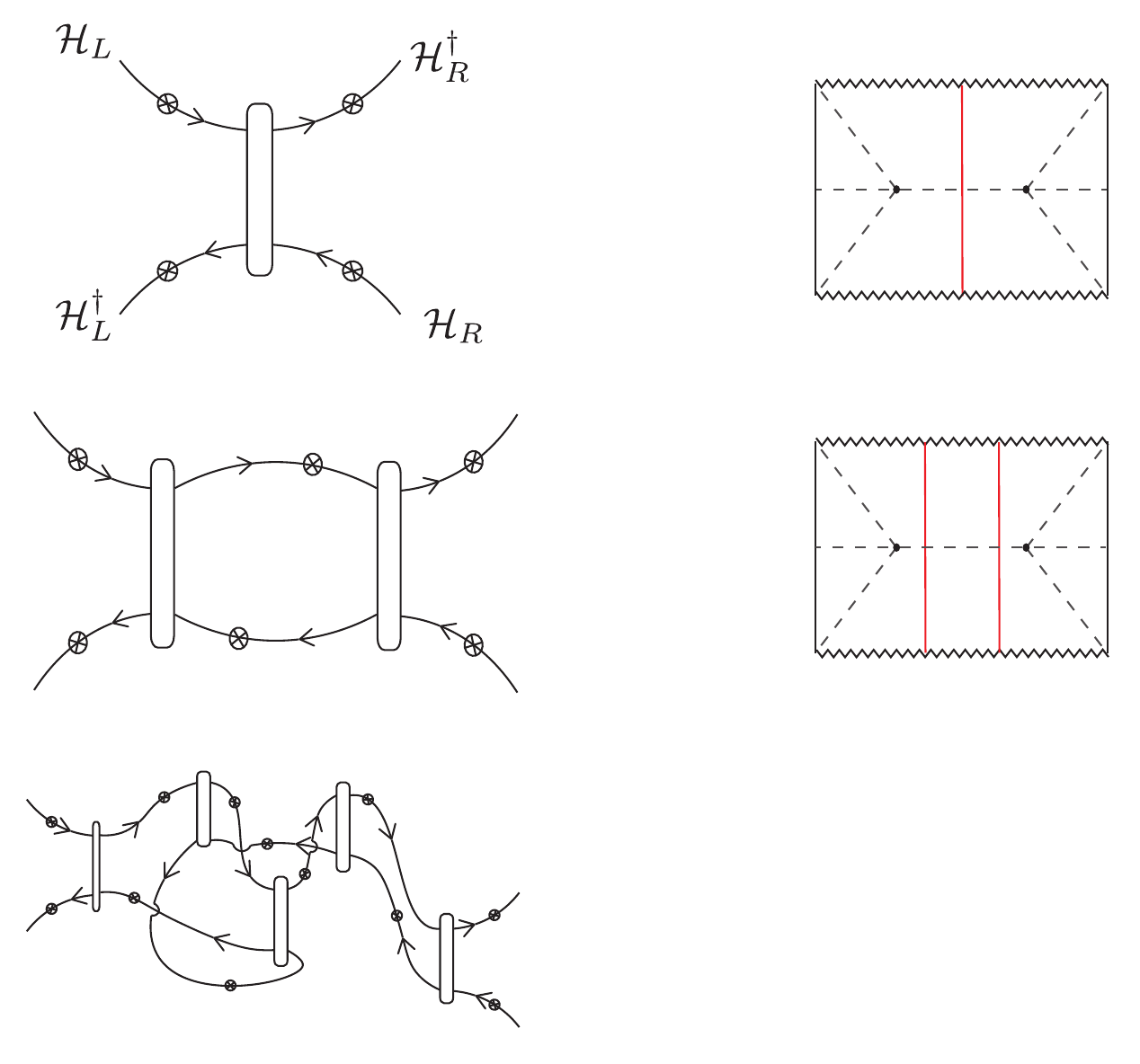}
    \caption{Single shot density matrix (first line), multishot density matrix (2nd line) and a more general wormhole (third line).}
    \label{fig:MultiShot}
\end{figure}

{\bf Multishot wormholes:} The single shot density matrix (\ref{eq:RandomSpinDM}) and its generalization  (\ref{eq:Non_Symm_dnsty_mtrxA}) can be graphically depicted as in the first line of figure \ref{fig:MultiShot}. The labels on the external lines encode which Hilbert space is associated with them. The $\otimes$ on the external legs indicate an action of some $e^{-\beta_i H}$, where $\beta_i$ can be complex. 

If we want to have two, more or less parallel, heavy particles generating a wormhole, we can use the expression in the second line of that figure (the arrangement of the external legs is the same). We can think about it as if we introduced another Hilbert space of states that live in the interior of the wormhole - let's call it I. We can then think about the density matrix as a single shot linking R to I and then another shot linking I to L. The two shots are invariant (under $SU(2)^N \rtimes S_N$ or under $SO(N)$) but they are linked by insertions of a Hamiltonian evolution, which is the only source of symmetry breaking. 

The latter is the simplest configuration. Clearly the most general transition matrix - and we conjecture, the most general semicalssical gravitational background that corresponds to a transition matrix - is given by a complicated network of shocks connected between themselves in different ways - for example the third line of that figure. In this case it describes 5 shots, some of which cross others. A general gas of particles in a wormhole will be described by a dense net of shots connected to each other in different ways, which encode the initial and final states of the wormhole. We are going to leave the study of these networks for future work. 

\subsection{The Majorana SYK model}\label{sec:MajSetUp}

So far we have been working on the random spin model introduced in section \ref{sec:RandomSpin}. Let us now consider the Majorana model of section \ref{sec:MajoranaModel} and discuss it in a parallel way.
In this section, we define a class of density matrices for the SYK model, which turns out to have the similar property as the ones we've discussed before. 

As in the random spin model, we are interested in density matrices invariant under the symmetry of the two-sided system. For the Majorana model, this symmetry is a diagonal $O(N)$ which rotates the left and right spinors simultaneously. Finding invariant density matrices is again equivalent to splitting $\mathcal H_L^\dagger \otimes \mathcal H_R$ into irreducible representations of this group. This is a cleaner representation theory problem than in the spin model case, and we have worked out the general invariant density matrices in appendix \ref{invdens}. We find there that a convenient basis for the space of invariant density matrices is provided by the 
\begin{equation} 
   G_M(A) =  \prod_i \left[1+iA \chi_L^i\chi_R^i\right]. 
\end{equation}
In the large $N$ limit, density matrices with different values of $A$ have disjoint support, and we take the density matrices $G_M(A)$ for different $A$ as our candidates for the dual of gravitational wormholes, analogous to the $G(A)$ in the random spin model defined in \eqref{SnglSht}. 

We can add a temperature by defining the density matrices
\begin{align}\label{eq:MajRho}
    G_M(A,\beta_L,\beta_R)=e^{-(\beta_RH_R+\beta_LH_L)/4}G_M(A)e^{-(\beta_RH_R+\beta_LH_L)/4}.
\end{align}

\subsubsection{Properties of $G_M(A)$}

\paragraph{Eigenvalues and entropy of $G_M(A)$}
In order to study the eigenvalues of 
\begin{align}
    G_M(A)\equiv \frac{1}{2^N}\prod_i \left[1+iA \chi_L^i\chi_R^i\right]
    \equiv \frac{1}{2^N}\prod_i G_{M,i}
\end{align}
we represent the Majorana fermions in terms of Pauli matrices.
The convenient representation for us is
\begin{align}
    \chi_L^{2j-1}&=
    \prod_{i=1}^{j-1}
    \left(\sigma^3_{i,L}\otimes \sigma^3_{i,R}\right) \otimes
    \left(-\sigma^1_{j,L}\otimes \mathbbm{1}_{j,R}\right)\otimes
    \prod_{i=j+1}^{N}
    \left(\mathbbm{1}_{i,L}\otimes \mathbbm{1}_{i,R}\right)\\ 
    \chi_L^{2j-1}&=
    \prod_{i=1}^{j-1}
    \left(\sigma^3_{i,L}\otimes \sigma^3_{i,R}\right) \otimes
    \left(\sigma^2_{j,L}\otimes \mathbbm{1}_{j,R}\right)\otimes
    \prod_{i=j+1}^{N}
    \left(\mathbbm{1}_{i,L}\otimes \mathbbm{1}_{i,R}\right)\\
    \chi_R^{2j-1}&=
    \prod_{i=1}^{j-1}
    \left(\sigma^3_{i,L}\otimes \sigma^3_{i,R}\right) \otimes
    \left(\sigma^3_{j,L}\otimes \sigma^2_{j,R}\right)\otimes
    \prod_{i=j+1}^{N}
    \left(\mathbbm{1}_{i,L}\otimes \mathbbm{1}_{i,R}\right)\\
    \chi_R^{2j}&=
    \prod_{i=1}^{j-1}
    \left(\sigma^3_{i,L}\otimes \sigma^3_{j,R}\right) \otimes
    \left(\sigma^3_{j,L}\otimes \sigma^1_{j,R}\right)\otimes
    \prod_{i=j+1}^{N}
    \left(\mathbbm{1}_{i,L}\otimes \mathbbm{1}_{i,R}\right)
\end{align}
% and 
% \begin{align}
% \begin{split}
%     \chi_L^3=\sigma^3\otimes \sigma^3\otimes (-\sigma^1)\otimes \mathbbm{1},\quad
%     \chi_L^4=\sigma^3\otimes \sigma^3\otimes\sigma^2\otimes \mathbbm{1},\\
%     \chi_R^3=\sigma^3\otimes \sigma^3\otimes\sigma^3\otimes \sigma^2,\quad 
%     \chi_R^4=\sigma^3\otimes \sigma^3\otimes\sigma^3\otimes \sigma^1
% \end{split}
% \end{align}
% and we continue this pattern to the rest of the Majorana fermions.
Here, $\sigma_{j,L}$ lives on the Hilbert space with the $(2j-1)$-th and $(2j)$-th left Majorana fermions, and likewise for the right Majorana fermions. 

% We call the small Hilbert space where the $2j$-th and $2j+1$-th Pauli matrices (in both left and right sides) live in as $\mathcal{H}_j$.

By using this representation, we have %{\bf MB What is j?}
\begin{align}
    G_M(A)=\prod_{j}\left(1-A\sigma^1_{j,L}\otimes \sigma^1_{j,R}-A\sigma^2_{j,L}\otimes \sigma^2_{j,R}-A^2\sigma^3_{j,L}\otimes \sigma^3_{j,R}\right).
\end{align}
This is the density matrix in the operator frame.

In the state frame (or in order to understand how $\rho(A)$ can be prepared using the Euclidean path integral) we repeat the Fierz identity exercise as in the Pauli matrices and write the density matrix as
%
% Let us now concentrate on the Hilbert space $\mathcal{H}_j$.
% This can be decomposed as $\mathcal{H}_j\equiv \mathcal{H}^j_L\otimes \mathcal{H}^j_R$, since we have two Pauli matrices to act on the Hilbert space.
% Even though $\chi_R^j$ acts on $\mathcal{H}_j^L$ as well, it does not trouble us, since everything comes in the combination of $\chi_L^j\chi_R^j$ and the action of $\chi_R^j$ merely changes the fermion parity on Hilbert space $\mathcal{H}_L^j$.
% We can also change the discussion to the Majorana fermions which commute between the two sides, by changing the definition of $\chi_R$ to $(-)^{F_L}\chi_R$.
%, as discussed earlier.
%
\begin{align}\label{eq:MajDnstA}
    G_M(A)&=\prod_{j}\left(\frac{(1+A)^2}{4}\mathbbm{1}_{j,I}\otimes \mathbbm{1}_{j,O}
    +\frac{1-A^2}{4}\left(\tau^1_{j,I}\otimes\tau^1_{j,O}+\tau^2_{j,I}\otimes\tau^2_{j,O}\right)+\frac{(1-A)^2}{4}\tau^3_{j,I}\otimes \tau^3_{j,O}\right),
\end{align}
where $\tau^a$ are defined the same way as in Sec. \ref{sec:SnglSpn}.

The entropy of this density matrix can also be computed, which is
\begin{align}
    S = -\frac{N}{2}\left[(1-A)\log\left(\frac{1-A}{2}\right)+(1+A)\log\left(\frac{1+A}{2}\right)\right].
\end{align}

\paragraph{Special values of $A$ and TFD states}

It is immediate from the eigenvalue decomposition of $\rho(A)$, the density matrix is pure at $A=\pm 1$.
This is in contrast to the random spin model, where $A=1$ corresponds to a pure state, while the other end of the spectrum, $A=-1/3$, was not a pure state, due to three states in the triplet.

Let us start from $A=1$.
This is, in the Euclidean preparation, having the unit operator inserted in the thermal half cycle.
This therefore corresponds to a TFD state at infinite temperature.
We can also think about it in the following way, that we have the relation
\begin{align}
    (\chi_L^i-i\chi_R^i)G_M(A=1)=0
\end{align}
and since $G_M(A=1)=\ket{\Psi_+}\bra{\Psi_+}$ is pure, this gives us the defining relation for the TFD state \cite{Maldacena:2018lmt},
\begin{align}
    (\chi_L^i-i\chi_R^i)\ket{\Psi_+}=0.
\end{align}

The case $A=-1$ is similar, but with a little twist.
This corresponds to inserting the fermion parity operator between the two half spaces when preparing the state (still as in figure \ref{fig:BulkRecon}),
and this has the effect of changing the definition of parity between the left and the right boundary.
This operator $G_M(A=-1)\equiv \ket{\Psi_-}\bra{\Psi_-}$ satisfies
\begin{align}
    (\chi_L^i+i\chi_R^i)G_M(A=-1)=0 \Longleftrightarrow (\chi_L^i+i\chi_R^i)\ket{\Psi_-}=0.
\end{align}
which is yet another choice of the TFD state given in \cite{Maldacena:2018lmt}.

The difference between $A=\pm 1$ is simply the difference in definitions of  the TFD state, if we twist the choice of the $CPT$ operator using the fermion number, in the definition of the TFD,
$\ket{\rm TFD}=\sum_n \ket{n}\otimes CPT\ket{n}$.

\paragraph{PETS interpretation}

Equation (\ref{eq:MajDnstA}) also gives us access to what operators we need to insert if we want to construct this space a-la PETS states \cite{Goel:2018ubv}. 
% 
% We can also define Majorana fermions $\psi$ \sfr{Is there a reason this is $\psi$ rather than $\chi$?}
% \mw{I thought it can be confusing since this Majorana is an operator from $H_L$ to $H_R$. Same reason as we renamed the Pauli as $\tau$. I don't have a strong opinion though.}
% as an operator from $\mathcal{H}_L$ to $\mathcal{H}_R$, whose convention we take as
% \begin{align}
%     \psi_{2j-1,(I,O)}=\tau^1_{j,(I,O)},\quad \psi_{2j,(I,O)}=\tau^2_{j,(I,O)},\quad (-)^{F_{j,(I,O)}}\equiv \psi_{2j-1,(I,O)}\psi_{2j+1,(I,O)}=\tau^3_{j,(I,O)}.
% \end{align}
% The density matrix then looks like
% \begin{align}
%     \rho(A)&=\prod_{j}\left(\frac{(1+A)^2}{4}
%     +\frac{1-A^2}{4}\left(\psi_{2j-1,I}\otimes\psi_{2j-1,O}+\psi_{2j,I}\otimes\psi_{2j,O}\right)+\frac{(1-A)^2}{4}(-)^{F_{j,I}}(-)^{F_{j,O}}\right),
% \end{align}
This means that the density matrix can be understood as inserting the operator $\mathbbm{1}$ with probability $\frac{(1+A)^2}{4}$, $\tau_1$ or $\tau_2$ with probability $\frac{1-A^2}{4}$, and the fermion parity operator with probability $\frac{(1-A)^2}{4}$.
For the density matrix to be unitary, we need $-1\leq A\leq 1$.

This is favourable, since the number of these operators inserted in the path integral follows the multinomial distribution, similar to the Pauli matrix case.
For example, the probability of finding operator $\psi$ for $K$ times follows the probability of
\begin{align}
    p(K)=\begin{pmatrix}
    N/2\\
    K
    \end{pmatrix}
    \left(\frac{1-A^2}{2}\right)^{K}\left(\frac{1+A^2}{2}\right)^{N/2-K}
\end{align}
and the distribution has a peak at large $N$.

\section{Explicit correlators in two-sided systems}\label{sec:CrlTwoSdd}

In this section we will evaluate the partition function and two sided correlation functions in our proposed (unnormalised) density matrices in the random spin model, i.e., 
\begin{align}
    \left\langle\Tr_{\mathcal{H}_L^\dagger \otimes \mathcal{H}_R}\left[G(A;\beta_L,\beta_R)\right]\right\rangle_J
    ,\ \ \left\langle\Tr_{\mathcal{H}_L^{\dagger} \otimes  \mathcal{H}_R }\left[G(A;\beta_L,\beta_R)\mathcal{O}_L\mathcal{O}_R\right]\right\rangle_J ,
\end{align}
with $G(A;\beta_L,\beta_R)$ defined in (\ref{eq:RandomSpinDM}). We would like, for example, to see how the length of the wormhole depends on $A$. We will set up the machinery mainly in the double scaling limit (and will also comment on the scaling of $A$ in the regular large $N$ limit). We will focus on the random spin model - the result of the Majorana SYK model is exactly parallel.

\subsection{Integrating over the random couplings} \label{sec:ChordComputations}

We will begin with the partition function
\begin{align}
    \left\langle\Tr_{\mathcal{H}_L^\dagger \otimes \mathcal{H}_R}\left[G(A;\beta_L,\beta_R)\right]\right\rangle_J
    =
    \left\langle\Tr_{\mathcal{H}_L^\dagger \otimes \mathcal{H}_R}\left[e^{-\beta_LH_L/4-\beta_RH_R/4}G(A)e^{-\beta_LH_L/4-\beta_RH_R/4}\right]\right\rangle_J.
\end{align}
By expanding everything as a power series in $H_{L,R}$, we will generally have to compute the moments
\begin{align} \label{eq:G(A)Moment}
    m_{k_L,k_R} \equiv \left\langle\Tr_{\mathcal{H}_L^\dagger \otimes \mathcal{H}_R}(G(A)H_L^{k_L}H_R^{k_R})\right\rangle_J .
\end{align}
In appendix \ref{sec:indices} we detail how $G$, $H_L$ and $H_R$ are contracted in order to implement the trace. The upshot is that we can rearrange the insertions of $H_L$ to bring it to the ordinary form of a matrix product where a lower index is contracted with an upper index of a matrix to the right.  We will denote this by $\Tr$ with no subscript. If we have some insertions of an additional operator then we need to keep track of their location. For example,
\begin{align}\label{eq:TwoPtCr}
\begin{split}
   & \left<\Tr_{{\cal H}_L^\dagger \otimes {\cal H}_R } \bigl( G(A,\beta_L,\beta_R) 
    O_R(t_R) O_L(t_L) \bigr)\right>_J \\
    &=\left<\Tr( e^{-\beta_L H_L/4} e^{-iH_L t_L} O_L(0) e^{iH_L t_L} e^{-\beta_L H_L/4} G(A) e^{-\beta_R H_R/4} e^{iH_Rt_R}  O_R(0) e^{-iH_Rt_R} e^{-\beta_R H_R/4} \bigr)\right>_J
\end{split}
\end{align}
where we defined $O_L(t_L)=e^{-iH_L t_L} O_L(0) e^{iH_L t_L}$ and $O_R(t_R)=e^{iH_R t_R} O_R(0) e^{-iH_R t_R}$ (for more details see appendix \ref{sec:indices}).

Next, we proceed as usual by using the fact that $\left\langle J_I J_{\tilde{I}}\right\rangle_J=\delta_{I,\tilde{I}}$. We then have 
\begin{align}
\begin{split}
    m_{k_L,k_R} &= \left\langle \Tr_{\mathcal{H}_L^\dagger \otimes \mathcal{H}_R}\left[G(A)H_R^{k_R}H_L^{k_L}\right]\right\rangle_J\\
    &= {3^{-pk/2}}\binom{N}{p}^{-k/2}
    \sum_{\text{pairings of $I$}}\sum_{\substack{I_1,\cdots,I_{k/2} \\ |I_i|=p}}
    \Tr_{\mathcal{H}_L^\dagger \otimes \mathcal{H}_R} 
    \left[ 
    G(A)
    \sigma_{R,I_1}\cdots \sigma_{R,I_{k_R}}
     \sigma_{L,I_{k_R+1}}\cdots \sigma_{L,I_{k}}
    \right]
    \\
    &= {3^{-pk/2}}\binom{N}{p}^{-k/2}
    \sum_{\text{pairings of $I$}}\sum_{\substack{I_1,\cdots,I_{k/2} \\ |I_i|=p}}
    \Tr
    \left[ 
    \sigma_{L,I_{k}}\cdots \sigma_{L,I_{k_R+1}}
    G(A)
    \sigma_{R,I_1}\cdots \sigma_{R,I_{k_R}}
    \right],
    \label{eq:cd}
\end{split}
\end{align}
where $k=k_L+k_R$. In the last equality we changed to an ordinary trace over $\mathcal{H}$, by flipping the order of operators in space $L$. 
We have written the sum as a sum over all the possible pairings, and over the composition of the index sets. We denote the set of all chord diagrams with $k$ nodes as $\text{CD}(k)$, such that the sum over $k$ pairs of $I$'s can be rewritten as $\sum_{\text{pairings of $I$}}=\sum_{\text{CD}(k)}$.

\begin{figure} [H]
    \centering
    \includegraphics[width=0.7\columnwidth,page=2]{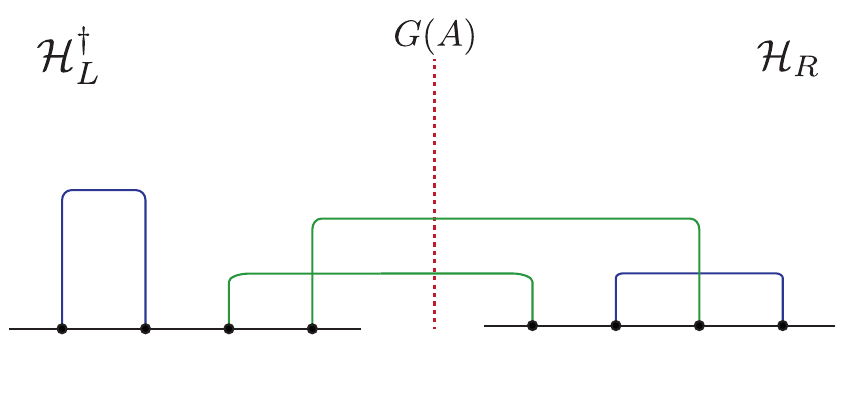}
    \caption{ Chord diagram description of 
        $\Tr_{\mathcal{H}_L^\dagger \otimes\mathcal{H}_R}
        \left(G(A)\sigma_{R,I_1}\sigma_{R,I_2}\sigma_{R,I_3}\sigma_{R,I_2}\sigma_{L,I_3}\sigma_{L,I_1}\sigma_{L,I_4}\sigma_{L,I_4}\right)=\Tr\left(\sigma_{L,I_4}\sigma_{L,I_4}\sigma_{L,I_1}\sigma_{L,I_3}G(A)\sigma_{R,I_1}\sigma_{R,I_2}\sigma_{R,I_3}\sigma_{R,I_2}\right)$,
    contributing to $m_{4,4}$. Note that the ordering of operators in space $L$ is with respect to the their ordering in the ordinary trace. This is the 3rd line of (\ref{eq:cd}).}
    \label{fig:CDinH2Dagger}
\end{figure}

We can now apply the chord technology as in appendix \ref{sec:DSLimitAndCD}: the $\sigma$ matrices on the right ``bubble" chords in $\mathcal H_R$. When we reach $G$, we are typically left with some chords open, i.e., an odd number of Pauli matrices in some of the species. We now need to take $G$ into account and carry out the trace on ${\cal H}_R$. Whenever there is an odd number of Pauli matrices in a species, we need to use the $A {\vec \sigma}_R{\vec \sigma}_L$ (for that species index) which will convert it into an incoming single $\sigma_L$. So the role of $G(A)$ is to convert\footnote{or, more precisely, pair}, in the species needed, an index from ${\cal H}_R$ to the same index in ${\cal H}_L^\dagger$. After $G(A)$ converts chords from space $R$ to space $L$  we then continue the bubbling process in space $L$. Of course various weights, associated with $A$, are incurred during this conversion - we will discuss them in a moment. 
% Similarly to appendix \ref{sec:DSLimitAndCD}, we can describe these computations in the language of chord diagrams. To do so, we first move from $\Tr_{\mathcal{H}_L^{\dagger} \otimes \mathcal{H}_R}$ to the ordinary trace description, by flipping the ordering of the operators in space $L$. We draw this by first placing the space $L$ operators, then $G(A)$, and only then the space $R$ operators, as in the third line of (\ref{eq:cd}). 
This is illustrated in figure \ref{fig:CDinH2Dagger}.

In this diagrammatic language, we can now present the different weights associated with chord intersections.

\paragraph{Weights of chord diagrams in the double scaling limit} \label{sec:2ChordCont}

As discussed in appendix \ref{sec:DSLimitAndCD} we can neglect the overlap of three index sets and treat each overlap of two index sets as independent. We can now start computing the contribution from two chords, and then multiply each contribution to evaluate a specific chord diagram.

Let $I_1,\ I_2,...$ denote the index sets associated to the different chords. We decompose the Hilbert space into site-wise components which we denote by the index $i$. Traces are always ordinary traces. \begin{itemize}
    \item {If $i$ does not appear in any of the chords}

The contribution from the Hilbert space ${\cal H}_i$ is simply
\begin{align}
    \Tr_i\left[G\left(A\right)\right] =1.
\end{align}

\item{If $i$ appears only once in just one chord - $I_1$}

In this case we have two possibilities.
If $I_1$ starts and ends in space $R$ (or likewise in space $L$), the contribution from Hilbert space ${\cal H}_i$ is
\begin{align}
    \frac{1}{3}\sum_{a}\Tr_i\left[G\left(A\right)\sigma_{R}^{a}\sigma_{R}^{a}\right] =1.
    \label{eq:1}
\end{align}
(The prefactor $1/3$ takes care of the prefactor in \eqref{eq:cd}.)
 
Meanwhile, if $I_1$ starts in space $L$ and ends in space $R$, we get
\begin{align}
    \frac{1}{3}\sum_{a}\Tr_i\left[\sigma_{L}^{a}G\left(A\right)\sigma_{R}^{a}\right]=A.
    \label{eq:A}
\end{align}

\item{If $i$ appears both in two chords - $I_1$ and $I_2$}

Similarly, we list all the possible contribution from Hilbert space ${\cal H}_i$ when $i$ appears both in $I_1$ and $I_2$:
\begin{align} \label{eq:FourPaulisTr}
&\begin{cases}
    %&
    \frac{1}{3^2}\sum_{a,b}\Tr_i\left[G\left(A\right)\sigma_{R}^{a}\sigma_{R}^{a}\sigma_{R}^{b}\sigma_{R}^{b}\right]=
    \frac{1}{3^2}\sum_{a,b}\Tr_i\left[\sigma_{L}^{a}\sigma_{L}^{a}\sigma_{L}^{b}\sigma_{L}^{b}G\left(A\right)\right]
    =1
    \\
    \frac{1}{3^2}\sum_{a,b}\Tr_i\left[\sigma_{L}^{b}\sigma_{L}^{b}G\left(A\right)\sigma_{R}^{a}\sigma_{R}^{a}\right]=1\\
    %&\qquad \qquad 
    \frac{1}{3^2}\sum_{a,b}\Tr_i\left[\sigma_{L}^{b}G\left(A\right)\sigma_{R}^{a}\sigma_{R}^{a}\sigma_{R}^{b}\right]=
    \frac{1}{3^2}\sum_{a,b}\Tr_i\left[\sigma_{L}^{b}\sigma_{L}^{b}\sigma_{L}^{a}G\left(A\right)\sigma_{R}^{a}\right]=A
\end{cases}\\
&\begin{cases}
    %&\qquad \qquad 
    \frac{1}{3^2}\sum_{a,b}\Tr_i\left[G\left(A\right)\sigma_{R}^{a}\sigma_{R}^{b}\sigma_{R}^{a}\sigma_{R}^{b}\right]=\frac{1}{3^2}\sum_{a,b}\Tr_i\left[\sigma_{L}^{a}\sigma_{L}^{b}\sigma_{L}^{a}\sigma_{L}^{b}G\left(A\right)\right]=-\frac{1}{3}
    \\
    %&\qquad \qquad 
    \frac{1}{3^2}\sum_{a,b}\Tr_i\left[\sigma_{L}^{b}G\left(A\right)\sigma_{R}^{a}\sigma_{R}^{b}\sigma_{R}^{a}\right]=\frac{1}{3^2}\sum_{a,b}\Tr_i\left[\sigma_{L}^{b}\sigma_{L}^{a}\sigma_{L}^{b}G\left(A\right)\sigma_{R}^{a}\right]=-\frac{A}{3}
    \\
    %&\qquad \qquad 
    \frac{1}{3^2}\sum_{a,b}\Tr_i\left[G\left(A\right)\sigma_{R}^{a}\sigma_{R}^{b}\sigma_{R}^{b}\sigma_{R}^{a}\right]=\frac{1}{3^2}\sum_{a,b}\Tr_i\left[\sigma_{L}^{a}\sigma_{L}^{b}\sigma_{L}^{b}\sigma_{L}^{a}G\left(A\right)\right]=1
\end{cases} \\
&\begin{cases}
    \frac{1}{3^2}\sum_{a,b}\Tr_i\left[\sigma_{L}^{a}G\left(A\right)\sigma_{R}^{a}\sigma_{R}^{b}\sigma_{R}^{b}\right]=\frac{1}{3^2}\sum_{a,b}\Tr_i\left[\sigma_{L}^{a}\sigma_{L}^{b}\sigma_{L}^{b}G\left(A\right)\sigma_{R}^{a}\right]=A
    \\
    %&\qquad \qquad \qquad \qquad 
    \frac{1}{3^2}\sum_{a,b}\Tr_i\left[\sigma_{L}^{a}\sigma_{L}^{b}G\left(A\right)\sigma_{R}^{a}\sigma_{R}^{b}\right]=\frac{1-2A}{3}
    \\
    %&\qquad \qquad \qquad \qquad 
    \frac{1}{3^2}\sum_{a,b}\Tr_i\left[\sigma_{L}^{b}\sigma_{L}^{a}G\left(A\right)\sigma_{R}^{a}\sigma_{R}^{b}\right]=\frac{1+2A}{3}
\end{cases}
\end{align}
\end{itemize}
In order to get the total contribution of the two chords, we can multiply contributions from each of the species's Hilbert spaces.
An index set contains $p$ indices, and so a chord running inside space $1$ or $2$ gets a factor of 1, and a chord running from space $1$ to $2$ gets a factor of $A^{p'}$, where $p'$ is the number of species that don't appear in any other chord, as one can see from \eqref{eq:1} and \eqref{eq:A}. 

In the double scaling limit $p'$ differs from $p$ by some additive finite amount (whereas they both scale like $\sqrt{N}$). 
We now also set $A=1-a/p$. Hence the final result in the double-scaling limit is
\begin{equation}
    B=A^{p'}=e^{-a}.
\end{equation}

This can conveniently be summarized in figures as
\begin{align}
     \raisebox{-.4\height}{\includegraphics[page=1,width=0.35\columnwidth]{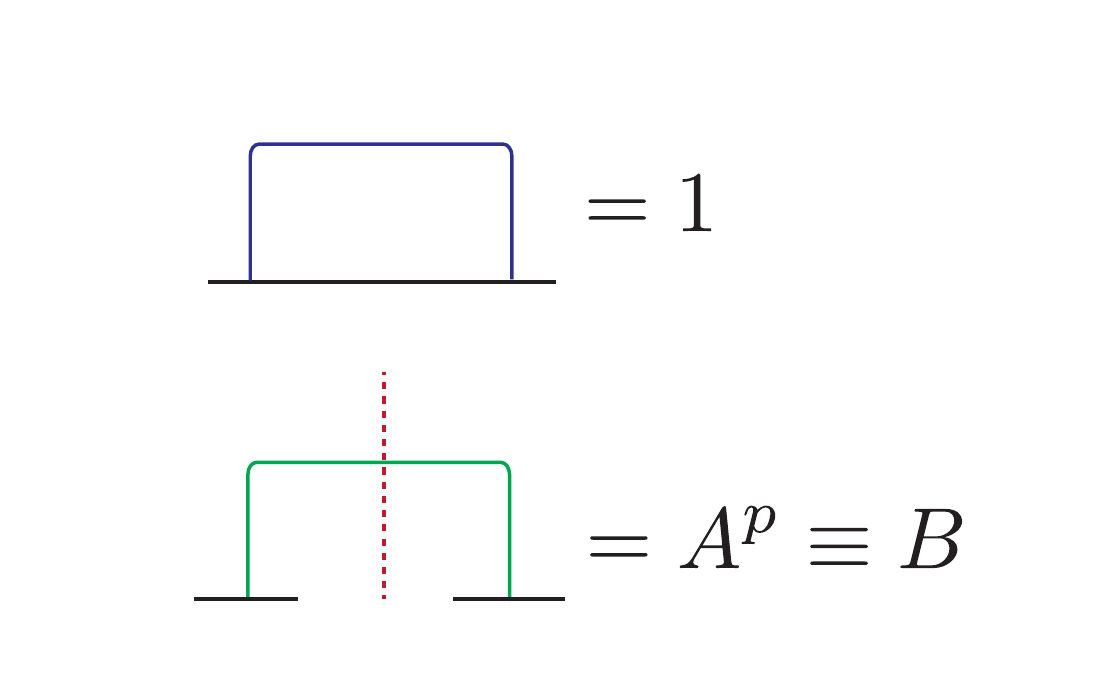}}
\end{align}
We depicted chords weighing $1$ in blue, while the ones weighing $A^p$ in green. The red line is there to remind us that we are computing the chord diagram in the presence of $G(A)$.

\begin{figure}[htbp]
\begin{center}
    \raisebox{-.4\height}{\includegraphics[page=2,width=0.6\columnwidth]{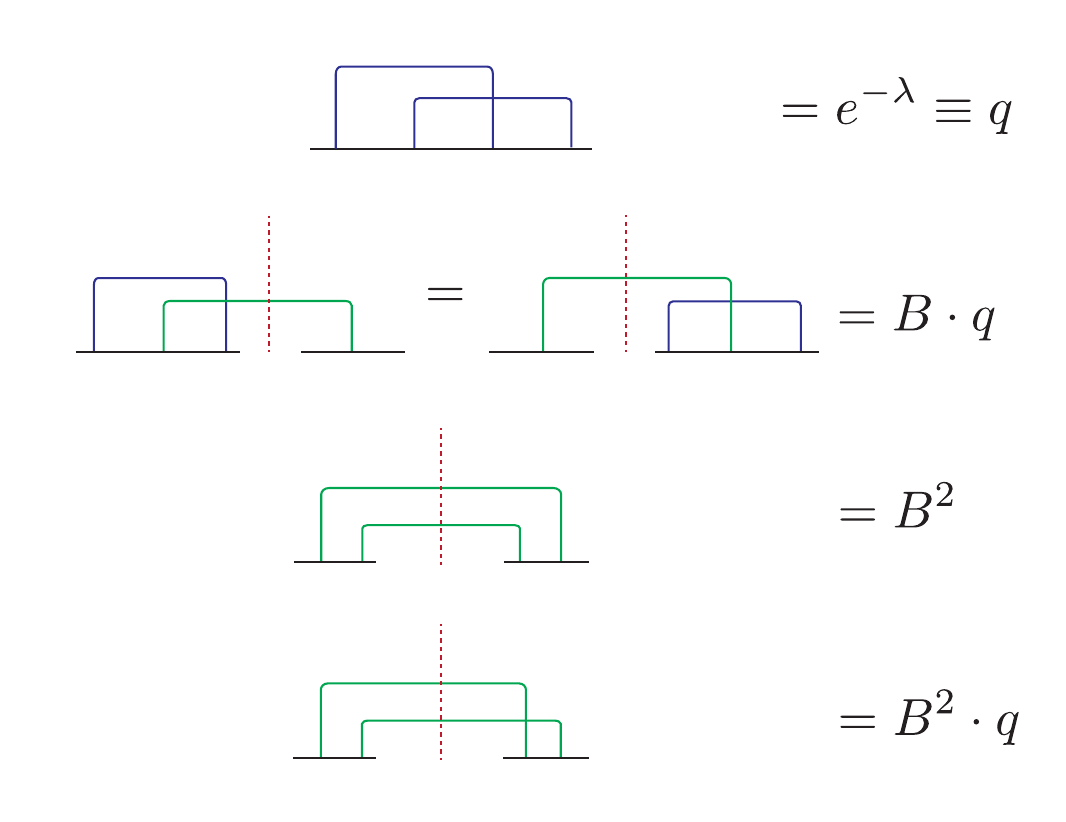}}
    \caption{Factors of chord configurations in the two space construction. We compute the relevant factors by multiplying the factors from (\ref{eq:FourPaulisTr}) by the number of times they appear, which follows a Poisson distribution, and setting $A=1-\frac{a}{p}$, as we're working in the double scaled limit. For example, the third diagram is given by
    \begin{minipage}{\linewidth}
    \begin{displaymath}
    e^{-p^2/N}\sum_{m=0}^{\infty}A^{2p-2m}\frac{(p^2/N)^m}{m!}\left(\frac{1+2A}{3}\right)^m=B^2e^{\frac{p^2}{N}(\frac{1+2A}{3A^2}-1)}=B^2,
    \end{displaymath} 
    \end{minipage}
    where we've used the fact that in the double scaled limit $p^2/N$ is finite, as well as $A=1-\frac{a}{p}$.
    }
    \label{fig:ChordFactors}
\end{center}
\end{figure}

Now that we assigned a value to each chord of two types above, we discuss the contribution coming from the intersections between two chords: as in appendix \ref{sec:DSLimitAndCD}, we can change variables in (\ref{eq:cd}) from $I_1,\cdots,I_{k/2}$ to the index overlaps $m_{ij}$, with $i,j=1,\cdots,k/2$, and change the measure accordingly. As discussed there, the measure is the Poisson distribution with parameter $p^2/N$, and the factor ${\binom{N}{p}}^{-k/2}$ ensures this distribution is properly normalized. This means that in order to find the contribution of a pair of chords, we take the relevant factor from (\ref{eq:FourPaulisTr}) for each overlapping index, and the total number of indices follows the Poisson distribution. The results are summarized in figure \ref{fig:ChordFactors}.

In conclusion the moment $m_{k_L,k_R}$ is given by
\begin{align} \label{eq:CPF}
    m_{k_L,k_R}=\sum_{\text{CD}(k_L,k_R)}q^{\text{\# intersections}}B^{\text{\# crossings}},
\end{align}
where $\text{CD}(k_L,k_R)$ are chord diagrams with 2 regions involving $k_L,k_R$ nodes respectively, and by $\text{crossings}$ we mean chords that cross from left to right.

\paragraph{Transfer matrix method} Next we would like to use the \textit{transfer matrix method} in order to compute the moment (\ref{eq:CPF}). To do so we use the same auxiliary Hilbert space and transfer matrix defined in appendix \ref{sec:DSLimitAndCD}. The similarity between (\ref{eq:ChordPF}) and (\ref{eq:CPF}) means that we need to do only small modifications to the moment computed by the transfer matrix (\ref{eq:TranMetMoment}). Indeed, the new ingredient is $B^{\# \text{crossings}}$, which can be accounted for simply by an operator insertion between the two regions. This means that we can compute the moment $m_{k_1,k_2}$ by
\begin{align} \label{eq:PartitionMoment}
    m_{k_L,k_R} = \left<0\right|T^{k_L}B^{\hat{N}}T^{k_R}\left|0\right>,
\end{align}
where
\begin{align}
    \hat{N}\ket{n}=n\ket{n}.
\end{align}
The insertion $B^{\hat{N}}$ is to account for the number of green chords, and $T$ is the transfer matrix defined in (\ref{eq:TransferMatrix}).

The moment (\ref{eq:PartitionMoment}) was previously computed in \cite{Berkooz:2018qkz}, where it arises in the computation of the two-point function of a random operator in a single copy of the system: if $M_B$ is a random operator of length $p'$, the two-point function involves
\begin{align}
    m_{k_L,k_R} = 2^{-N}\left<\Tr_{\mathcal{H}}(H^{k_L}M_BH^{k_R}M_B)\right>_{J,\tilde{J}},
\end{align}
which was shown in \cite{Berkooz:2018qkz} to be given by (\ref{eq:PartitionMoment}) with $p'=-\frac{p\log B}{\lambda}$. This is in line with our expectation that the partition function of the two sided space is given by an insertion and extraction of a massive operator when taking the Euclidean boundary to be a circle (as in figure \ref{fig:BulkRecon}).

This correspondence can be generalized to include higher point functions. For example one just takes the 2nd line in equation \eqref{eq:TwoPtCr} and converts it into a chord prescription
\begin{equation}
    \langle 0| e^{-(\beta_L/4-it_L)T} O_L e^{(-\beta_L/4+it_L)T} B^{\hat N} e^{(-\beta_R/4-it_R)T} O_R e^{(-\beta_R/4+it_R)T}|0\rangle 
\end{equation}
and we take $O_L=O_R$, an Hermitian random operator of length ${\hat p}$, which is replaced by a chord whose index set is of this length. 
%Also, we defined $O_L(t_L)\equiv $
This can also be written as a four-point function in a single SYK copy
\begin{equation}
2^{-N}\left\langle \text{Tr}_{\mathcal{H}}\left(M_{B}e^{-\frac{\beta_{L}}{4}H}O_L\left(t_{L}\right)e^{-\frac{\beta_{L}}{4}H}M_{B}e^{-\frac{\beta_{R}}{4}H}O_R\left(t_{R}\right)e^{-\frac{\beta_{R}}{4}H}\right)\right\rangle _{J,J',\tilde{J}}.
\end{equation}
We will borrow the formulas for this expression from \cite{Berkooz:2018jqr}.

In a similar manner, the $n$-point function evaluated in a doubled system over the density matrix $G(A)$ can be mapped to $n+2$ point functions evaluated in a single copy.

\subsection{Size of correlators and length of wormholes}\label{sec:LenghWH}

We would like to get some sense of the relation between the length of the wormhole and the parameter $A$ appearing in $G(A,\beta_L,
\beta_R)$ in (\ref{eq:RandomSpinDM}). The computation was basically done in section \ref{sec:PropSnglSht}, but we have just now derived all the factors in a more systematic way. 

Consider a two-sided two-point function of an operator $\mathcal{O}$ made from a sum over strings of ${\tilde p}$ Pauli matrices with random coefficients as defined in (\ref{eq:OpDef}). Given that the operator is made out of ${\tilde p}$ symbols, its IR conformal dimension is ${\tilde p}/p$. We will place this operator in the Left and Right Hilbert spaces at times $t_R=t_L=0$. First take  $\beta_L=\beta_R=0$. Then as in the discussion in the previous section, each Pauli matrix that gets converted from Left to Right by $G(A)$ gives a weighting by $A$, so the conversion of an operator with $\tilde p$ Pauli matrices gives 
\begin{align}\label{2PTA}
    \Tr[G(A)\mathcal{O}_L\mathcal{O}_R]=A^{\tilde p}=B^{{\tilde p}/p} .
\end{align}

Adding temperature to the system (by going to $\rho(A,\beta_L,\beta_R))$ will slightly modify the calculation, but we still need to have that the chord of the probe operator goes from Right to Left, hence we will still have a contribution like  \eqref{2PTA}. There will be a modified prefactor that depends on the $\beta$'s, but the ${\tilde p}$ scaling of the RHS of  \eqref{2PTA} and the qualitative dependence on $B$ will remain similar. 

On the gravity side, the two-point function behaves as $e^{-mL}$, where $L$ is the length of the wormhole, and for heavy operators the mass $m$ is related to the conformal dimension by $m = \Delta/R_{AdS} = \tilde p/p R_{AdS}$. Hence we have 
\begin{align}
    e^{-m L}\propto B^{{\tilde p}/p} \rightarrow L/R_{AdS} \propto -\log(B)=a
\end{align}
When $A=1$, then the wormhole is of length 0 as we expect, and the wormhole becomes longer as we move away from the thermofield double.   
 
This is consistent with the observation above for the entropy of $\rho(A)$ that as $A$ moves away from 1 the entropy of the density matrix increases - as the length of the wormhole increases, then more states can be accommodated in the wormhole (with the same gravitational profile).
 
The formulas given above are for the double scaling limit. In the finite $p$ case, one might worry that the situation is different because $B=A^p$ is finite for any value of $A$ and not just $A$ close to one ($p$ is a fixed integer and need not be particularly large). We would like to argue however that a semiclassical wormhole still exists only for $A$ close to 1 (as some power of $1/N$). The reason is that for finite $p$, one obtains  semiclassical gravity by going to low temperatures, which means a high power of $H$ if we use the moment method. In fact one has to take the number of $H$'s to scale like a positive power of $N$. To have a semiclassical wormhole we similarly require a large number of chords going between the right and left spaces. I.e, the number of chords that go from left to right scale like $n_{LR}\sim N^\alpha,$ for some $\alpha>0$. In order to have a finite $B^{n_{LR}}$ we need again to take $A=1-a/N^\alpha$.

\subsection{Partition function and two point functions}

By using the transfer matrix and the expression for the moments given above, we can compute the partition function and the $2$-point functions of the two-sided random spin system.
Because of the discussion above, we can just cite the result given in \cite{Berkooz:2018qkz} to get the final result.

\paragraph{Partition function}

For the \textit{partition function}, the result is
\begin{align}
\begin{split}
    &\left<\Tr_{\mathcal{H}_L^\dagger \otimes\mathcal{H}_R}\left[G(A;\beta_L,\beta_R)\right]\right>_J=
    2^{-N}\left<\Tr_{\mathcal{H}}\left[e^{-\frac{\beta_L}{2}H} M_B e^{-\frac{\beta_R}{2}H}M_B\right]\right>_{J,\tilde{J}}\\
    =&\int_0^{\pi}\prod_{i=1}^2\left\{\frac{d\theta_i}{2\pi}\left(q,e^{\pm 2i\theta_i};q\right)_\infty
    \exp\left(-\frac{\beta_jE(\theta_j)}{2}\right)
    \right\}
    \frac{\left(B^2;q\right)_\infty}{\left(Be^{ i(\pm\theta_1\pm\theta_2)};q\right)_\infty}.
\end{split}
\end{align}
where $E(\theta_j) = \frac{2\cos\theta_j}{\sqrt{1-q}}$, $(a;q)_\infty$ are q-Pochammer symbols, defined by
\begin{align}
\begin{split}
    (a;q)_n&\equiv\prod_{k=0}^{n-1}(1-aq^k),\  (a;q)_\infty\equiv\prod_{k=0}^{\infty}(1-aq^k)\\
    (a_1,a_2,\dots,a_k;q)_n&\equiv \prod_{i=1}^k (a_k;q)_n\ , (e^{\pm i\theta};q)\equiv (e^{+ i\theta};q)(e^{- i\theta};q)
\end{split}
\end{align}
and further details are given in appendices \ref{sec:DSLimitAndCD} and \ref{sec:SpecialFunctions}.

For later convenience, we will define a new quantity $\ell_B$ by
\begin{align}
    B\equiv q^{\ell_B}
\end{align}
This corresponds to the conformal dimension of the operator $M_B$.

\paragraph{Double-sided two-point function}

For the \textit{double-sided two-point functions} we can cite the results for the crossed four-point function
\begin{align}
    \begin{split}
    &\left<\Tr_{\mathcal{H}_L^\dagger \otimes\mathcal{H}_R}\left[G(A;\beta_L,\beta_R)M_L(t_L)M_R(t_R)\right]\right>_{J,J'}
    \\
    & =\left\langle \text{Tr}_{\mathcal{H}}\left(e^{-\frac{\beta_{L}}{4}H}M\left(t_L\right)e^{-\frac{\beta_{L}}{4}H}M_{B}e^{-\frac{\beta_{R}}{4}H}M\left(t_{R}\right)e^{-\frac{\beta_{R}}{4}H}M_{B}\right)\right\rangle _{J,J',\tilde{J}}
    \\
    &= \int _0^{\pi }  \prod _{j=1} ^4 \left\{ \frac{d\theta _j}{2\pi } (q,e^{\pm 2i\theta _j} ;q)_{\infty } \exp{\left(- \gamma_j E(\theta_j) \right)} \right\}
    \frac{\left(B e^{-i(\theta_2+\theta_3)}, \tilde{q} B e^{i(\theta_3\pm \theta_1)}, \tilde{q} B e^{i(\theta_2\pm \theta_4)} ;q \right)_{\infty } }{\left(\tilde{q}^2 B e^{i( \theta_2+\theta_3)} ;q \right)_{\infty } } \\
    & \qquad \times  \frac{\left(\tilde{q}^2,\tilde{q}^2, B^2;q \right)_{\infty } }{\left(B e^{i(\pm  \theta _{2} \pm \theta _{3} )}, B e^{i(\pm  \theta _{1} \pm \theta _{4} )}, \tilde{q} e^{i(\pm  \theta _{1} \pm \theta _{2} )}, \tilde{q} e^{i(\pm  \theta _{3} \pm \theta _{4} )} ;q \right)_{\infty } } \\
& \qquad \times  {}_8W_7\left( \frac{\tilde{q}^2 B e^{i(\theta_2+\theta_3)} }{q}; B e^{i(\theta_2+\theta _3)} ,\tilde{q} e^{i(\theta _2 \pm \theta _1)}, \tilde{q} e^{i(\theta _3 \pm \theta _4)} ;q, B e^{-i(\theta_2+\theta _3)} \right),
    \end{split}\label{eq:DoubleSided2pt}
\end{align}
with $\gamma_1 = \frac{\beta_R}{4}+it_R,\gamma_3=\frac{\beta_L}{4}+it_L$ and $\gamma_2=\gamma_1^*,\gamma_4=\gamma_3^*$. Remember that $\tilde{q}=e^{-2\frac{p\tilde{p}}{N}}$, as defined in (\ref{eq:q_tilde}). The basic hypergeometric series $_8W_7$ is defined in (\ref{eq:DEF:8W7Series}).
We can also check that it reduces to the partition function when we take $\ell_M\to 0$.

\subsection{Low energy limit}

The formulas above are quite general, and can be generalized further to any correlator across the wormhole, and in fact, using a discussion similar to \ref{sec:TrnsMtrc}, can be generalized to particles entering from the past singularity or leaving through the future singularity. Here we will have a more modest goal of just testing our machinery using the two sided correlator in the shock wave \cite{Shenker:2013pqa}. The shock wave is a particularly convenient background as the two-sided correlation function was evaluated for any length/shockwave strength, and it plays a key role in the study of quantum chaos in black holes. 

We will first review some formulas that have to do with the low energy limit and then we will specialize them to the shock wave in the next subsection.

\subsubsection{Useful formulae} \label{sec:LowEnergyFormulas}

The low-energy limit we take is the same as in \cite{Berkooz:2018jqr}, where one takes $q\to 1$ and the low-energy limit at the same time in a consistent way, and is known to reproduce the result of the Schwarzian quantum mechanics, or equivalently, of JT gravity.
We summarize below all the necessary formulae without derivation, for which the reader is referred to \cite{Berkooz:2018jqr} and references therein.

First take $q=e^{-\lambda}$ and take $\lambda\to 0$.
At the same time, take
\begin{align} \label{eq:Def:y_var}
    \theta_j\equiv \pi-\lambda y_j
\end{align}
so that
\begin{align}
    E(\theta_j)\equiv \frac{2\cos\theta_j}{\sqrt{1-q}}=E(\theta=\pi)+\lambda^{\frac{3}{2}}y_j^2
\end{align}
This $E(\theta)$ is the energy eigenvalue of the SYK model in the double-scaling limit, as discussed in the previous sections.
In the following, for the sake of simplicity, we shift $E(\theta=\pi)$ to $0$.

Also, in the $\lambda\to 0$ limit, we have
\begin{align}
    \int_0^{\pi}\frac{d\theta_i}{2\pi}\left(q,e^{\pm 2i\theta_i};q\right)_\infty
    =\frac{\lambda(q;q)_{\infty}^3(1-q)^2}{2\pi^2}\int_{0}^{\infty}d\left(y^2\right)
    \sinh(2\pi y)
\end{align}
and
\begin{align}
    \frac{\left(B^2;q\right)_\infty}{\left(Be^{ i(\pm\theta_1\pm\theta_2)};q\right)_\infty}
    =\frac{(1-q)^{-3+2\ell_B}}{(q;q)_\infty^3}\cdot \frac{\Gamma(\ell_B\pm iy_1\pm iy_2)}{\Gamma(2\ell_B)}.
\end{align}
where $B=q^{\ell_B}$.
This $\ell_B$ can be understood as the operator dimension of the fictitious operator $M_B$ introduced in the previous subsection.
In particular, in the double-scaling limit $A=1-\frac{a}{p}$, we have
\begin{align}
    B=e^{-a}\equiv q^{\ell_B}=e^{-\lambda \ell_B} \Longleftrightarrow \ell_B=\frac{a}{\lambda}.
\end{align}
Here $a=O(p^0)$, but can still depend on $\lambda$ non-trivially.
We have also defined $\ell_M$ by using $\tilde{q}\equiv q^{\ell_M}$, and it is the dimension of the random operator, whose two-point function we compute later. 

We will eventually use the saddle-point approximation to evaluate the forthcoming integrals --
the consistency of the saddle-point approximation is that we are in the low temperature limit, which is
\begin{align}
    \lambda^{-\frac{1}{2}}\ll \beta \ll \lambda^{-\frac{3}{2}}
\end{align}
In principle, this should be checked every time one does the saddle-point approximation, but we refer the reader to \cite{Berkooz:2018qkz} and proceed.

\subsubsection{The low-energy limit and the shock wave}
\label{sec:tpf}

\paragraph{Double-sided two point function}

We first compute the low energy limit of the unnormalised double-sided two point function.
We begin with expression  \eqref{eq:DoubleSided2pt}, for the two sided correlator with R and L insertions at ${\tilde t}_{L,R}+t_d$. We will eventually take $t_d$ to infinity, but we shift the dependence on $t_d$ into a time evolution of $M_B$. This shifts the entry and exit point of the defect in the past and future singularities. 
In other words, we have the unnormalised density matrix
\begin{align} \label{eq:time_dep_rho_2_side}
    \rho \equiv e^{-(\beta_L+4it_d) H_L/4-(\beta_R-4it_d) H_R/4}G(A)e^{-(\beta_L-4it_d) H_L/4-(\beta_R+4it_d) H_R/4}.
\end{align}

The expression that we get is therefore (with a small cyclic rearrangement of the terms)
\begin{align} \label{eq:2SidedCorrel}
\begin{split}
    &\langle \Tr\left[\rho M_L(t_L)M_R(t_R)\right]\rangle
    =\langle \Tr\left[e^{-\frac{\beta_R}{4}H}M(t_R)e^{-\frac{\beta_R}{4}H}M_B(-t_d)e^{-\frac{\beta_L}{4}H}M(t_L)e^{-\frac{\beta_L}{4}H}M_B(-t_d)\right]\rangle\\
    =&
    \langle \Tr\left[e^{-\left(\frac{\beta_R}{4}-i(t_R+t_d)\right)H}Me^{-\left(\frac{\beta_R}{4}+i(t_R+t_d)\right)H}M_Be^{-\left(\frac{\beta_L}{4}-i(t_L+t_d)\right)H}Me^{-\left(\frac{\beta_L}{4}+i(t_L+t_d)\right)H}M_B\right]\rangle
\end{split}
\end{align}

We now take the low energy limit using the formulas presented above.
\begin{align}
\begin{split}
    &\langle \Tr\left[\rho M_L(t_L)M_R(t_R)\right]\rangle
    =\left(\frac{\lambda}{2\pi^2}\right)^4(q;q)^3_\infty (1-q)^{-1+2\ell_M+2\ell_B}\\
    &\quad\! \times \int\prod_{j=1}^{4}\left\{dy_j^2 \sinh(2\pi y_j^2)\right\}\\
    &\quad \times
    \exp\left(-\lambda^{\frac{3}{2}}\beta M
    -\frac{\beta_L\lambda^{\frac{3}{2}}}{2}\omega_B-
    \left(\frac{\beta_R}{4}+i(t_R+t_d)\right)\lambda^{\frac{3}{2}}\omega_1-\left(-\frac{\beta_L}{4}+i(t_L+t_d)\right)\lambda^{\frac{3}{2}}\omega_2\right)\\
    &\quad \times 
    \frac{\Gamma(\ell_B-i(y_3\pm y_2))\Gamma(\ell_B+i(y_1\pm y_4))}{\Gamma(2\ell_B)}
    \frac{\Gamma(\ell_M+i(y_3 \pm y_4))\Gamma(\ell_M-i(y_1\pm y_2))}{\Gamma(2\ell_M)}\\
    &\quad \times\int_{-i\infty}^{i\infty}\frac{du}{2\pi i}
    q^{u-\ell_M+i(y_3-y_4)}
    \frac{\Gamma(-u+\ell_B+i(y_4-y_1))\Gamma(-u+\ell_M+i(y_4-y_3))}{\Gamma(u+\ell_B-i(y_4-y_1))\Gamma(u+\ell_M-i(y_4-y_3))}\\
    &\quad \quad \times \Gamma(u)\Gamma(u-i(y_2+y_4-y_1-y_3))\Gamma(u-2iy_4)\Gamma(u+i(y_1+y_2+y_3-y_4))
\end{split}
\end{align}
where
\begin{align}
    y_1^2\equiv M+\omega_B,\quad y_2^2=M+\omega_B-\omega_2,\quad y_3^2=M+\omega_1,\quad y^2_4=M.
\end{align}
This is equivalent to \cite{Lam:2018pvp}. As in the single-sided case, we can a posteriori justify that $\omega_{1,2,B}$ are much smaller than $M$ (for further justification see \cite{Berkooz:2018qkz}).
The integral over $u$ can be done using the contour integral -- at large $M$, it can been shown that the only relevant poles are at $u=0$ and at $u=i(y_2+y_4-y_1-y_3)$ \cite{Lam:2018pvp}. More precisely, the other poles are suppressed for any $t_d$, and between these two poles, one of them is dominant depending on whether $t_d$ goes to $\infty$ or $-\infty$. This corresponds to the two possible orientations of the shockwave. 

We first analyze the $u=0$ pole.
\begin{align}
\begin{split}
    &\Tr\left[\rho_0M_R(t_R)M_L(t_L)\right]\biggl|_{u=0\text{ pole}}
    =\left(\frac{\lambda}{2\pi^2}\right)^4(q;q)^3_\infty (1-q)^{-1+2\ell_M+2\ell_B}\\
    &\quad\! \times \int
    dM\,d\nu_B\, d\nu_1\, d\nu_2\,
    (4M)^{\ell_M+\ell_B}
    \exp\left(2\pi \sqrt{M}-\lambda^{\frac{3}{2}}\beta M+\pi \nu_B\left(\frac{\beta_R-\beta_L}{\beta_L+\beta_R}\right)\right)\\
    &\quad \times\exp\left({ -\pi\nu_1\left(\frac{1}{2}\frac{\beta_R-\beta_L}{\beta_L+\beta_R}+\frac{2i(t_R+t_d)}{\beta}\right)+\pi\nu_2\left(\frac{1}{2}\frac{\beta_R-\beta_L}{\beta_L+\beta_R}-\frac{2i(t_L+t_d)}{\beta}\right)}\right)\\
    &\quad \times \left(2\sqrt{M}\right)^{i(\nu_1+\nu_2)}
    e^{\lambda\ell_M-\frac{2\pi i \sqrt{\lambda}}{\beta}\nu_1}\\
    &\quad \times \frac{\Gamma(\ell_B-i\nu_B)\Gamma(\ell_B+i(\nu_B-\nu_1-\nu_2))\Gamma(\ell_M-i\nu_1)\Gamma(\ell_M-i\nu_2)\Gamma(i(\nu_1+\nu_2))}{\Gamma(2\ell_M)\Gamma(2\ell_B)}
\end{split}
\end{align}
where as always $\nu_{B,1,2}\equiv\frac{\omega_{B,1,2}}{2\sqrt{M}}$.

The $\nu_B$ integral can be performed by using $\ell_B,\nu_B\gg \nu_1,\nu_2$ at the saddle point and then using the contour integration,
\begin{align}
    \int_{-\infty}^{\infty}\frac{d\nu_B}{2\pi}\, e^{\pi\nu_B\left(\frac{\beta_R-\beta_L}{\beta_L+\beta_R}\right)}\frac{\Gamma(\ell_B- i\nu_B)\Gamma(\ell_B+i(\nu_B-\nu_1-\nu_2))}{\Gamma(2\ell_B)}
    =\left(\frac{1}{2\cos \left(\frac{\pi}{2}\frac{\beta_R-\beta_L}{\beta_L+\beta_R}\right)}\right)^{2\ell_B}\ell_B^{-i(\nu_1+\nu_2)}.
\end{align}
We get
\begin{align}
\begin{split}
    &\Tr\left[\rho_0M_R(t_R)M_L(t_L)\right]\biggl|_{u=0\text{ pole}}
    \\
    &=\left(\frac{\lambda}{2\pi^2}\right)^4(q;q)^3_\infty (1-q)^{-1+2\ell_M+2\ell_B}\int
    dM\,d\nu_1\, d\nu_2\,
    (4M)^{\ell_M}M^{\ell_B}
    \exp\left(2\pi \sqrt{M}-\lambda^{\frac{3}{2}}\beta M\right)\\
    &\quad \times\exp\left(-\frac{2\pi i}{\beta}(t_R\nu_1+t_L\nu_2)\right)
    \\
    &\quad \times \frac{\Gamma(\ell_M-i\nu_1)\Gamma(\ell_M-i\nu_2)\Gamma(i(\nu_1+\nu_2))}{\Gamma(2\ell_M)}\times \left(\frac{\ell_B}{2\sqrt{M}}e^{\frac{2\pi t_d}{\beta}}\right)^{-i(\nu_1+\nu_2)}\left(\frac{1}{2\cos \left(\frac{\pi}{2}\frac{\beta_R-\beta_L}{\beta_L+\beta_R}\right)}\right)^{2\ell_B}.
\end{split}
\end{align}
The $u=i(y_2+y_4-y_1-y_3)$ pole can likewise be evaluated. It is much smaller then $u=0$ when we take $t_d\rightarrow\infty$ (it is the dominant pole when $t_d\rightarrow -\infty$). 

We can now do the $M$ integral, which is localised at the saddle-point
\begin{align}
    \sqrt{M_0}\equiv \frac{\pi\lambda^{-\frac{3}{2}}}{\beta}
    =\pi\lambda^{-\frac{3}{2}}T,
\end{align}
and the final expression becomes
\begin{align}
\begin{split}
    &\Tr\left[\rho_0M_L(t_L)M_R(t_R)\right]
    =C\int dE_L\,dE_R\,
    e^{-it_LE_L-it_RE_R}
    \\
    &\quad \times \frac{\Gamma\left(\ell_M-\frac{iE_L}{2\pi T}\right)\Gamma\left(\ell_M-\frac{iE_R}{2\pi T}\right)\Gamma\left(\frac{i(E_L+E_R)}{2\pi T}\right)}{\Gamma(2\ell_M)}\times \left(\frac{\lambda^{\frac{3}{2}}\ell_B}{2\pi T}e^{2\pi T t_d}\right)^{\frac{-i(E_L+E_R)}{2\pi T}},
\end{split}
\end{align}
where
\begin{align}
    E_L\equiv \lambda^{\frac{3}{2}}\omega_2, \quad E_R\equiv \lambda^{\frac{3}{2}}\omega_1
\end{align}
and
\begin{align}
    C\equiv \left(\frac{\lambda}{2\pi^2}\right)^4(q;q)^3_\infty (1-q)^{-1+2\ell_M+2\ell_B}(4M_0)^{\ell_M}M^{\ell_B}
    \exp\left(\pi^2\lambda^{-3/2}T\right)
    \left(\frac{1}{2\cos \left(\frac{\pi}{2}\frac{\beta_R-\beta_L}{\beta_L+\beta_R}\right)}\right)^{2\ell_B}
\end{align}

Finally, we need to normalize by the partition function. Since we get the partition function by taking $\ell_M=0$, and hence taking the saddle where $E_1=E_2=0$ in the integral in the above expression, we simply get
\begin{align}
    \Tr[\rho]=C.
\end{align}

It therefore follows that the normalised two-point function in the low-energy limit becomes
\begin{align}
\begin{split}
    &\langle M_L(t_L)M_R(t_R)\rangle
    =\int dE_L\,dE_R\,
    e^{-it_LE_L-it_RE_R}
    \\
    &\quad \times \frac{\Gamma\left(\ell_M-\frac{iE_L}{2\pi T}\right)\Gamma\left(\ell_M-\frac{iE_R}{2\pi T}\right)\Gamma\left(\frac{i(E_L+E_R)}{2\pi T}\right)}{\Gamma(2\ell_M)}\times \left(\frac{\lambda^{\frac{3}{2}}\ell_B}{2\pi T}e^{2\pi T t_d}\right)^{\frac{-i(E_L+E_R)}{2\pi T}},
\end{split}\label{eq:QMtwopoint}
\end{align}
which should be used in the next subsection, where we compare with the gravity computation.

A comment is in order.
Our result indicates that the normalised two-point function does not depend on the difference between $\beta_1$ and $\beta_2$.
This is natural, because we demanded that $\ell_M$ is much smaller than $\ell_B$.
In the dual gravity picture, changing $\beta_1-\beta_2$ corresponds to changing the location of the end-of-the-worldline created by a particle of mass $\ell_M$ \cite{Lam:2018pvp}.
In the approximation we are using, this is too light to affect the two-point function that we are computing.
The effect will be visible at subleading order in the saddle-point approximation, since the cancellation of $\left({2\cos \left(\frac{\pi}{2}\frac{\beta_R-\beta_L}{\beta_L+\beta_R}\right)}\right)^{-2\ell_B}$ between the partition function and the unnormalised two-point function is due to the approximation $\ell_M,\nu_B\gg \nu_M$ in the $\nu_B$ integral.

\subsubsection{Comparison to JT gravity}

In the previous section we started from a particle going from the past region of the black hole to the future region. This is slightly different than the construction in \cite{Shenker:2013pqa} where one throws in the particle from the spatial (field theory) boundary. Still, in the limit $t_d\rightarrow\infty$ the particle grazes the horizon and we expect it to converge to the shock wave geometry. 

The two-point function that we would like to compare is determined by the renormalised geodesic distance $d$ of two boundary points in the shockwave geometry,
\begin{align} \label{eq:GeodesicApprox}
    \langle M_L(t_L)M_R(t_R)\rangle\sim e^{-m_{\rm probe} d},
\end{align}
where $m_{\rm probe}$ is the mass of a probe particle.
Since JT gravity is an s-wave reduction of the three-dimensional Einstein gravity \cite{Maxfield:2020ale}, we can use the result from \cite{Shenker:2013pqa}, which is
\begin{align}
    e^{-m_{\rm probe} d}=\left(\cosh\left(\pi T_{\rm BH}\left(\tilde{t}_{R}-\tilde{t}_{L}\right)\right)+\frac{\alpha_{\rm shock}}{2}e^{-\pi T_{\rm BH}(\tilde{t}_{R}+\tilde{t}_{L})}\right)^{-2m_{\rm probe}\ell_{\rm AdS}}.
\end{align}
where the operators are inserted at times ${\tilde t}_{L,R}$ on the two sides, and $T_{BH}=R_{BH}/2\pi l_{AdS}^2 $. $\alpha_{\text{shock}}$ quantifies the strength of the shock and is given by $\alpha_{\text{shock}}=(E/M_{BH}) e^{2\pi T_{BH}{\tilde t}_w}$ where ${\tilde t}_w$ is the time of shock and we take it to infinity as we take $E/M_{BH}\rightarrow 0$ keeping $\alpha_{\text{shock}}$ fixed\footnote{The computation was done in the theory of three-dimensional gravity, but since they considered the shockwave to be S-wave, we can do the  dimensional reduction and then the result is the same as in JT gravity.}. 
%We have switched here to the conventions of  \cite{Shenker:2013pqa} where time flows downwards in left, and we will keep track that we are using this convention by using the symbols ${\tilde t}$. So ${\tilde t_R}=t_R,\ {\tilde t_L}=-t_L$ {\bf MB Do we want to do this?}
%\mw{It's just convention, but I agree that it's obnoxious. We can change the QM computation, but we will have to invert the exponent of the time evolution and I didn't have enough power to do all of them. But I think this QM definition is the remnant of doing $H_1\otimes H_2$ notation, so this should be cleared away in the end for sure.... If had taken $H_1\otimes H_2\dagger$, the natural flow of time would have matched with that of gravity.}
By Fourier transforming in terms of $\tilde{t}_{L,R}$, this can be rewritten as
\begin{align}
\begin{split}
    e^{-m_{\rm probe}d}
    =
    \frac{1}{(2\pi T_{\rm BH})^2}&\int \frac{d\tilde\omega_L}{2\pi}\,\frac{d\tilde\omega_R}{2\pi}\,
    e^{-i\tilde{t}_L\tilde\omega_L-i\tilde{t}_R\tilde\omega_R}
    \left(\alpha_{\rm shockwave}\right)^{-\frac{i(\tilde\omega_L+\tilde\omega_R)}{2\pi T_{\rm BH}}}
    \\
    &\quad\times
    \frac{\Gamma\left(m_{\rm probe}\ell_{\rm AdS}-\frac{i\tilde\omega_L}{2\pi T_{\rm BH}}\right)\Gamma\left(m_{\rm probe}\ell_{\rm AdS}-\frac{i\tilde\omega_R}{2\pi T_{\rm BH}}\right)\Gamma\left(\frac{i(\tilde\omega_L+\tilde\omega_R)}{2\pi T_{\rm BH}}\right)}{\Gamma(2m_{\rm probe}\ell_{\rm AdS})}
\end{split}
\end{align}

This expression is of the same form as \eqref{eq:QMtwopoint}, which supports the claim that the density matrix \eqref{eq:time_dep_rho_2_side} is dual to the shockwave geomtry in JT gravity.
The parameters between the two are matched
by
\begin{align}
\begin{split}
    \tilde{\omega}_{L,R}&\Longleftrightarrow E_{L,R}\\
    \tilde{t}_{L,R,w}&\Longleftrightarrow t_{L,R,d}\\
    m_{\rm probe}\ell_{\rm AdS}&\Longleftrightarrow \ell_M\\
    T_{\rm BH}&\Longleftrightarrow T\\
    \alpha_{\rm shockwave}&\Longleftrightarrow \frac{\lambda^{\frac{3}{2}}\ell_B}{2\pi T}e^{2\pi T t_d}
\end{split}
\end{align}

% with the correspondence summarised in Table \ref{table:match}.
% {\bf MB Turn this into inline}

% \begin{table}[htbp]
% \begin{center}
%   \begin{tabular}{c|c|c}
%     SYK model & JT gravity & Comment \\\hline\hline
%     $\lambda^{\frac{3}{2}}t_{L,R}$ & $\displaystyle-\frac{\tilde{t}_{L,R}}{\ell_{\rm AdS}}$ & \specialcell{The definition of the coordinate system flips the sign.}\\\hline
%     $\lambda^{\frac{3}{2}}t_d$ & $\displaystyle\frac{\tilde{t}_{w}}{\ell_{\rm AdS}}$ & \specialcell{The point of insertion of $M_B$ was at $-t_d$.}\\\hline
%     $\lambda^{-\frac{3}{2}}T$ & $\displaystyle-\frac{T_{\rm BH}}{\ell_{\rm AdS}}$ & \\\hline
%     $\ell_M$ & $m_{\rm probe}\ell_{\rm AdS}$ & \\\hline
%     $\displaystyle\frac{\ell_B}{2\pi\lambda^{-\frac{3}{2}}T}e^{2\pi T_{\rm} t_d}$ & $\alpha_{\rm shockwave}$ & This means $\displaystyle\frac{\ell_B}{2\pi\lambda^{-\frac{3}{2}}T}=\frac{E_{\rm shockwave}}{4M_{\rm BH}}$.\\\hline
%   \end{tabular}
%   \caption{
%  }\label{table:match}
% \end{center}
% \end{table}

\section{Summary and Outlook}

We have studied generalizations of the thermofield double state, looking for suitable density matrices on a doubled Hilbert space that could correspond to smooth semi-classical wormholes in a dual gravitational description. We argued that in generalising the thermofield double state, it is natural to consider density matrices on the doubled Hilbert space, averaging over microscopic details that do not affect the gravitational profile in the bulk dual. Studying density matrices also makes it possible for us to translate from a state frame description to the operator frame description, which allowed us to formulate a criterion for the existence of a wormhole in the dual description in terms of the operator conversion properties in the operator frame description. 

We have considered these questions explicitly in the context of the SYK model and a related random spin model. These models have a diagonal symmetry after averaging over the random couplings -- $O(N)$ for the SYK model and $SU(2)^N \rtimes S_N$ for the random spin model -- and we argued that we should consider density matrices invariant under these symmetries up to explicit insertions of the Hamiltonian. We focused our considerations on a one-parameter family of invariant density matrices in each model, \eqref{eq:RandomSpinDM} for the random spin model and \eqref{eq:MajRho} for the SYK model. 

The minimal length of the wormhole is simply encoded in the parameter of the density matrix (and can also be simply related to its entropy). The ER bridge is encoded by correlations of the random terms of the Hamiltonians on the two sides, and the length is encoded by the extent to which correlations are suppressed.   

For the random spin model, we calculated the two-sided correlation function for probe operators in this density matrix directly in the spin system, and found a nice correspondence with previous gravitational calculations of shock wave deformations of the eternal black hole in a low energy limit. The result is actually the same for the Majorana once we reduce the latter to the same chord prescription, as in \cite{Berkooz:2018jqr}. This provides a proof of principle calculation demonstrating that our proposed density matrices have the expected properties to correspond to semiclassical wormholes. 

There are number of potential directions for further development: 
\begin{itemize}
    \item It would be interesting to further study the properties of our density matrices and understand their dual gravitational description in more detail. 
    \item The density matrices we considered are  ``single shot'': they correspond to a single localised perturbation in the wormhole in the bulk. It would be interesting to extend our study to the ``multi shot'' transition matrices introduced in section \ref{sec:TrnsMtrc} and relate to previous discussion of wormholes with multiple shocks in the bulk.
    \item From the point of view of the invariance property, there are many possible generalizations of the simple density matrices we considered. The most general invariant density matrices probably do not have a simple gravitational description, but there may be other classes which do; it would be valuable to explore this space further. 
    \item These density matrices correspond to wormholes with two boundaries. It would be interesting to also understand the general description of multi-boundary wormholes \cite{Maxfield:2014kra,Balasubramanian:2014hda,Marolf:2015vma}, in particular if we could shed some further light on the nature of the entanglement structure in the dual of these geometries. In the SYK or spin models, it is straightforward to set up density matrices on a Hilbert space with $k$ factors, but it is less clear in this case what are the natural density matrices to consider. Our operator conversion and invariance discussions do not have a straightforward extension. 
\end{itemize}

\section*{Acknowledgements} 
It is our pleasure to thank Ofer Aharony, Ping Gao, and Ho Tat Lam for useful discussions. 
The work of SFR is supported by STFC through grant ST/T000708/1. The work of MB, NB and MW is supported by Israel Science Foundation center for excellence grant (grant number 2289/18), and by the German Research Foundation through a German-Israeli Project Cooperation (DIP) grant "Holography and the Swampland".
The work of MW is supported by
the Foreign Postdoctoral Fellowship Program of the Israel Academy of Sciences and Humanities. MB is
the incumbent of the Charles and David Wolfson Professorial Chair of Theoretical Physics.

\appendix

\section{Some comments on indices and the matching of operators between right and left}\label{sec:indices}

Purifying the state with ${\cal H}^\dagger$ is a bit non-standard so we would like to go over the associated index conventions.

\medskip
    
{\bf 1.} Suppose we choose some basis in ${\cal H}$ - let us call it $e_a$. The vector of coefficients is then some $\psi^a$. The dual basis in ${\cal H}^\dagger$ is $e^a$, and the vector of coefficients in ${\cal H}^\dagger$ is then $w_a$. With these convention, the matrices in  \eqref{eq:PauliChnls} have the following index structure
$ {(\tau_I)}^{a_R}_{b_L},\ {(\tau_O)}^{a_L}_{b_R},\ {(\sigma_R)}^{a_R}_{b_R},\ {(\sigma_L)}^{a_L}_{b_L}$. The $\tau$ matrices used in \eqref{eq:defsinglet} are $\tau_I$'s.

For example, an operator ${\cal H}_R\rightarrow {\cal H}_R$, which is an object in ${\cal H}_R\otimes {\cal H}_R^\dagger$, is given by \begin{equation}
    O^a_{\ b}e_{1,a}\otimes {e_1^\dagger}^b \in {\cal H}_R\otimes {\cal H}_R^\dagger\ ,\ \ \psi^a e_a \rightarrow (O^a_{\ b}\psi^b) e_a\ .
\end{equation}
% % For these matrices, the upper index is a row index and the lower index is the column index in the standard matrix notation. 
% If needed we will indent the column indices to the right, but it's easier just to remember that when action on ${\cal H}$ are concerned we contract the lower index with a vector index on the right. 
An operator from ${\cal H}^\dagger$ to itself is given by
\begin{equation}
    U^{\ a}_{b} e^{\dagger,b}\otimes e_{a}\in {\cal H}^\dagger\otimes {\cal H},\ \ {w'}_{b}= U^{\ a}_{b} w_{a}
\end{equation}
However, both $O$ and $U$ are really defined in the same class of objects and can be identified. In short, an operator on ${\cal H}$ can be used to define an operator on ${\cal H}^\dagger$ when we act with the other index on the vector of coefficients\footnote{If we have a preferred basis and use it to define anti-unitary transformation which identifies vectors in ${\cal H}$ and ${\cal H}^\dagger$, then the definition above is the transpose in that basis - but we do not need to use such a map in our conventions.}.  

\medskip

{\bf 2.} Returning to our density matrices, $G$ now carries the indices 
\begin{equation}
    G^{a_Ra_L}_{b_Rb_L}
\end{equation}
where indices $b_L, a_R$ are indices of a vector of coefficient in ${{\cal H}_L^\dagger \otimes \cal H}_R$, and $a_L, b_R$ are the indices in the dual space. 
This means that if we have several operators $O_i:{\cal H}_R\rightarrow {\cal H}_R$ and several operator $U_j: {\cal H}_L^\dagger\rightarrow {\cal H}_L^\dagger$ then the trace which computes $\langle O_1...O_l U_1...U_{l'}\rangle_G$ is
\begin{equation}
G^{a_Ra_L}_{b_Rb_L}{(O_1)}^{b_R}_{c_{R,1}}\dots{(O_l)}^{c_{R,l-1}}_{a_R}{(U_1)}_{a_L}^{c_{L,1}}... {(U_{l'})}_{c_{L,l'-1}}^{b_L}
\end{equation}
% This is the required index structure because upper and lower indices for us are related to whether its an ${\cal }$ or ${\cal H}^\dagger$ space, and indenting an index forward means that it's a column index. As before we can forget about indenting indices and just remember that in space 2 an operator acts on a vector in this space by contracting the upper index in $U$ with the vector index. 
% What one needs to be careful with is which index is contracted when we discuss an operator acting on a vector in ${\cal H}_1$, a vector in ${\cal H}_2^\dagger$ or their tensor product. 
% With these convention suppose we are given a bunch of operators $O_1,...,O_K$ in space R and a bunch of operators $U_1,..., U_l$ on space 2 \nb{should probably write $U:\mathcal{H}_2^{\dagger}\to\mathcal{H}_2^{\dagger}$}, then the expectation value in the density matrix G is given by
% \begin{equation}\label{eq:HHDagTrace}
%     \Tr(GO_1\cdots O_l U_1\cdots U_l)=G^{a_1a_2}_{b_1b_2} 
%     (O_1)^{b_1}_{\ c_1} (O_2)^{c_1}_{\ d_1}... (O_l)^{z_1}_{\ a_1}
%     (U_1)_{a_2}^{\ c_2} (U_2)_{c_2}^{\ d_2}... (U_l)_{z_2}^{\ b_2}
% \end{equation}

\medskip

{\bf 3.} When we compute a two sided correlator of some Hermitian operator, then we are given such an operator on ${\cal H}$ and we know what to insert in the right Hilbert space. As we discussed before, an operator on ${\cal H}_R$ can be used just as well to define an operator on the left side Hilbert space ${\cal H}_L^\dagger$
\begin{equation}\label{eq:OpId}
U^a_b=O^a_b.
\end{equation}
So suppose we are given some Hermitian operator on ${\cal H}_R$. In an AdS/CFT context, in the bulk it would correspond to a real field. Now take the field and move it to the left side. The operator that we get is the operator just defined.

For example consider the Hamiltonian on ${\cal H}_R$, and it has a series of eigenvectors and eigenvalues 
\begin{equation}
    H^a_b \psi_i^b = E_i \psi^a
\end{equation}
Then the Hamiltonian as it acts on ${\cal H}_L^\dagger$ is
\begin{equation}
    w^i_a H^a_b = \lambda_i w^i_b
\end{equation}
where $w_i$ is the dual basis of $\psi_i$  
\begin{equation}
    w^i_a \psi_j^a = \delta^i_j
\end{equation}
So the prescription above indeed keeps the eigenvalues of the Hamiltonian (or any other Hermitian operator).  

\medskip

{\bf 4.}
Time evolution for ${\cal H}^\dagger$ looks a bit unusual in these conventions
\begin{align}
\begin{split}
    &States:\ \psi^a(t)= \bigl( e^{-iHt} \bigr)^a_b \psi^b(0),\ \ \ w_b(t)= w_a(0)   \bigl( e^{-iHt} \bigr)^a_b\\
    &Operators:\ (O_1(t))^a_b= (e^{iHt})^a_c (O(0))^c_d (e^{-iHt})^d_b,\ \ \   (U_2(t))^a_b= (e^{iHt})^c_b (U_2(0))_c^d (e^{-iHt})^a_d\\
\end{split}
\end{align}
Note that the identification \eqref{eq:OpId} is such that it identifies $O^a_b(t)=U^a_b(-t)$ as expected from the overall invariance of the background when $A=1$.

\section{Motivating restriction to $G(A)$}
\label{ent}

In our discussion of the Pauli spin model, we focused on the simple invariant density matrix introduced in \eqref{SnglSht}, which is simply the product of invariant density matrices for each of the spins with the same value of $A$. In this appendix, we consider generalising this to allow different values of $A$ for different spins; this gives an $SU(2)^N$ invariant density matrix
\begin{equation} \label{suninvt}
\rho = \prod_{i=1}^N \frac{(1 +  A_i \vec{\sigma}^i \cdot \vec{\sigma}^i)}{4}.   
\end{equation}
We can obtain $SU(2)^N \rtimes S_N$ invariance by summing over all permutations of the indices with equal weight. So we consider
\begin{equation} \label{pauliinvt} 
\rho = \frac{1}{|S_N|}\sum_{p \in S_N} \prod_{i=1}^N (1 +  A_{p(i)} \vec{\sigma}^{i} \cdot \vec{\sigma}^i).
\end{equation}
This gives us an $N$ parameter family of $SU(2)^N \rtimes S_N$ invariant density matrices. Writing \eqref{pauliinvt} more explicitly in terms of states, it is 
\begin{equation} \label{pistates}
    \rho = \frac{1}{|S_N|}\sum_{p \in S_N} \prod_{i=1}^N \left( (1- \hat A_{p(i)}) |s\rangle \langle s| + \frac{\hat A_{p(i)}}{3} \sum_m |m,t \rangle \langle m,t| \right).  
\end{equation}
Taking the product, there are ${\binom{N}{K}} 3^K$ states with $K$ of the modes in triplet states. All these states have the same probability $p_K$, 
\begin{equation}
  p_K = \frac{1}{3^K |S_N|} \sum_{p \in S_N} \prod_{i = 1}^{N-K} (1- \hat A_{p(i)}) \prod_{i = N-K+1}^N \hat A_{p(i)}. 
\end{equation}

We now argue that setting the $A_i$ equal maximises the entropy of the density matrix, for fixed strength of the correlation between the two boundaries, providing some additional motivation for this choice.

For simplicity, we will give an explicit calculation where we consider just two possible values for $\hat A_i$, that is there are two values $\hat A$ and $\hat B$ which appear in $N_A,N_B$ of the species ($N_A+N_B=N$). We define $\alpha_{A,B} = N_{A,B}/N$. It should be straightforward, although probably messy calculationally, to extend the argument to more general cases.

We want to consider varying $\hat A- \hat B$ holding fixed the ``length of the wormhole" in the dual description. We take the size of double-sided two-point correlations as a proxy for this length, as in the bulk, these are related to the length of geodesics between the two boundaries, as seen in (\ref{eq:GeodesicApprox}). Let $M_i$ be a random boundary operator of length $\tilde p$, according to the definition in (\ref{eq:OpDef}). Consider now the double sided 2-pt function $\left<M_1(t_1)M_2(t_2)\right>$. On the SYK side we can use appendix \ref{sec:DSLimitAndCD} to see that in the double scaled limit
\begin{align}
    \left<M_1(0)M_2(t)\right> \propto A^{\alpha_A \tilde p}B^{\alpha_B \tilde p}.
\end{align}
Thus, our condition is that we vary $\hat A - \hat B$ holding fixed 
\begin{align}
    W = A^{\alpha_A}B^{\alpha_B}=(A/B)^{\alpha_A}B=\left(\frac{3/4-\hat{A}}{3/4-\hat{B}}\right)^{\alpha_A}\left(1-\frac{4}{3}\hat{B}\right).
\end{align}

We consider a density matrix
\begin{equation}
    \rho = \frac{1}{|P_N|}\sum_{p \in P_N} \left( (1- \hat A) |s\rangle \langle s| + \frac{\hat A}{3} \sum_m |m,t \rangle \langle m,t| \right)^{N_A} \left( (1- \hat B) |s\rangle \langle s| + \frac{\hat B}{3} \sum_m |m,t \rangle \langle m,t| \right)^{N_B}.  
\end{equation}
There are ${\binom{N}{K}} 3^K$ states with $(N-K)$ states in the singlet and $K$ states in the triplet. After summing over permutations, all these states enter the density matrix $\rho$ with probability
$p_K = \tilde p_K F(z)$, where
\begin{equation}
   \tilde p_K = \frac{{\binom{N_B}{K}}}{{\binom{N}{K}} 3^K}  (1- \hat A)^{N_A} (1 - \hat B)^{N_B-K} \hat B^{K}, \quad z = \frac{A(1-B)}{B(1-A)} , \quad F(z) = {}_2 F_1(-K,-N_A, 1-K+N_B; z). 
\end{equation}
The entropy for the density matrix is 
\begin{equation}
    S = - \sum_{K=0}^N {\binom{N}{K}} 3^K p_K \ln p_K.
\end{equation}

We consider varying $\hat A$, $\hat B$, holding $W = A^{N_A} B^{N_B}$ fixed. Let's consider the variation around some common value $\hat A = \hat B = \hat C$. At linear order, $\hat A = \hat C + \alpha_A^{-1} \gamma$, $\hat B = \hat C - \alpha_B^{-1} \gamma$, and the first derivative of the entropy is 
\begin{equation}
    \frac{\partial S}{\partial \gamma} = \frac{1}{\alpha_A} \frac{\partial S}{\partial \hat A} - \frac{1}{\alpha_B} \frac{\partial S}{\partial \hat B} = - \sum_{K=0}^N {\binom{N}{K}} 3^K ( 1 + \ln p_K)  ( \frac{1}{\alpha_A} \frac{\partial p_K}{\partial \hat A} - \frac{1}{\alpha_B} \frac{\partial p_K}{\partial \hat B} ).
\end{equation}
We have
\begin{equation}
     \frac{\partial p_K}{\partial \hat A} = \frac{\partial \tilde p_K}{\partial \hat A} F(z) + p_K \frac{\partial z}{\partial \hat A} \frac{dF}{dz}, \quad \frac{\partial p_K}{\partial \hat B} = \frac{\partial \tilde p_K}{\partial \hat B} F(z) + p_K \frac{\partial z}{\partial \hat B} \frac{dF}{dz},
\end{equation}
and $\frac{dF}{dz}(z=1) = \frac{K N_A}{N} F(z=1)$. This gives, at $\gamma=0$, 
\begin{equation}
 \alpha_A^{-1} \frac{\partial p_K}{\partial \hat A} =   \alpha_B^{-1}  \frac{\partial p_K}{\partial \hat B}  = \left( - \frac{N-K}{(1- \hat C)} + \frac{K}{\hat C} \right) p_K.
\end{equation}
Thus, $\alpha_A^{-1}  \frac{\partial p_K}{\partial \hat A} - \alpha_B^{-1} \frac{\partial p_K}{\partial \hat B} = 0$ at $\gamma=0$, and hence $\frac{\partial S}{\partial \gamma} =0$; $\gamma=0$ is an extremum of the entropy. 

To see whether it's a minimum or a maximum, we need to consider the second derivative. First, work out the dependence on $\gamma$ to second order: if $\hat A = \hat C + \alpha_A^{-1} \gamma + a \gamma^2$, $\hat B = \hat C - \alpha_B^{-1} \gamma + b \gamma^2$, the variation will have constant $A^{N_A} B^{N_B}$ if 
\begin{equation}
    \alpha_A a + \alpha_B b = -\frac{4}{6 C} \alpha_A^{-1} \alpha_B^{-1}. 
\end{equation}
Now we correct our previous formula for the first derivative to 
\begin{equation}
    \frac{\partial S}{\partial \gamma} = (\alpha_A^{-1} + 2a \gamma) \frac{\partial S}{\partial \hat A}  + (- \alpha_B^{-1} + 2 b \gamma)  \frac{\partial S}{\partial \hat B},
\end{equation}
so 
\begin{equation}
    \frac{\partial^2 S}{\partial \gamma^2}|_{\gamma=0} = 2a \frac{\partial S}{\partial \hat A} + 2 b  \frac{\partial S}{\partial \hat B} + \alpha_A^{-2} \frac{\partial^2 S}{\partial \hat A^2} - 2 \alpha_A^{-1} \alpha_B^{-1}  \frac{\partial^2 S}{\partial \hat A \partial \hat B} + \alpha_B^{-2}  \frac{\partial^2 S}{\partial \hat B^2}. 
\end{equation}
After some calculation, we find
\begin{equation}
    \frac{\partial^2 S}{\partial \gamma^2}|_{\gamma=0} = - 2(N_A a+ N_B b)  \ln \left( \frac{\hat C}{3 (1- \hat C)} \right)=  \frac{8 N \alpha_A^{-1} \alpha_B^{-1} }{6(1- \frac{4}{3} \hat C)}   \ln \left( \frac{\hat C}{3 (1- \hat C)} \right).
\end{equation}
This gives a negative second derivative, indicating a maximum.

This gives a motivation for preferring the simple density matrix \eqref{SnglSht}. We should note however that the most general $SU(2)^N \rtimes S_N$ invariant density matrix is more complicated than \eqref{pauliinvt}. In the basis of singlet and triplet states, the general density invariant density matrix is block diagonal, as in \eqref{pistates}, but the states with $K$ triplets do not form an irreducible representation of the symmetry, so the most general density matrix is not proportional to the identity on the space of states with $K$ triplets. The irreducible representations of $S_N$ on the space of states with $K$ triplets are labelled by Young tableaux with up to two rows, and the second row has up to $L$ boxes, where $L = \mathrm{min}(K, N-K)$. So there are $L+1$ irreps in the space of states with $K$ triplets, and the general $SU(2)^N \rtimes S_N$ invariant density matrix is
\begin{equation} \label{dtrips} 
\rho = \sum_{K=0}^N \sum_{I= 0}^L p_{K,I} \rho_{K,I}, 
\end{equation}
where $\rho_{K,I}$ are the projectors onto the irreps. For $N$ even, this has $\frac{1}{4} (N+2)^2$ parameters $p_{K,I}$, subject to the overall constraint  $\sum_K \sum_I p_{K,I} =1$.\footnote{For $N$ odd, it's $\frac{1}{4} (N+1)(N+3)$.}  We will not explore this more general family of invariant density matrices here; we see no reason to expect these more refined objects have a nice geometrical dual.

\section{$O(N)$ invariant density matrices in the Majorana model}
\label{invdens}

We here discuss the symmetry-invariant density matrices for the Majorana model. The effective symmetry for the Majorana SYK model is $O(N)$. This is the symmetry of rotating the index of Majorana fermions in each space at the same time. One may also require that the density matrix be symmetric under the exchange of spaces $R$ and $L$.

\paragraph{Representation structure} Let us now briefly remind the reader the representation structure of $O(N)$ (we will use the notations of \cite{Polchinski:1998rr}). We can write the total space as a tensor product of the left and right spaces. Restricting for concreteness to even $N$, then each Hilbert space is the sum of the two spin representations of $O(N)$ with different chiralities. We will refer to them as $S$ and $S'$. So in total the doubled Hilbert space  is $(S\oplus S')_R\otimes (S\oplus S')_L$.

There is one word of caution, however.
Since $\chi_L$ and $\chi_R$ anticommutes rather than commutes, $\chi_R$ also acts and flips the chirality of the left Hilbert space.
Instead, $(-)^{F_L}\chi_R$ commutes with $\chi_L$ and hence acts trivially on the left Hilbert space.

Next consider an $SO(N)$-invariant density matrix on this space. First recall that
\begin{align}\label{RepProd}
\begin{split}
    S\otimes S&=[0]_L+[2]_L+...[N/2]_+\\
    S'\otimes S'&=[0]_R+[2]_R+...[N/2]_-\\
    S\otimes S'&=[1]_L+[3]_L+..+[N/2-1]_L\\
    S'\otimes S&=[1]_R+[3]_R+..+[N/2-1]_R\\
\end{split}
\end{align}
where $[..]$ are the antisymmetric tensor representations and $[N/2]_\pm$ is the (anti) self dual tensor rep. See that under $SO(N)$ there is no meaning to the subindex $L/R$, and the representations are equivalent. This means that
\begin{equation}
    \left({\cal H}_L^\dagger\otimes {\cal H}_R \right)_{SO(N)} = 2[0]+2[1]+..+2[N/2-1]+[N/2]_++[N/2]_-
\end{equation}
so overall there are $(N/2+2)$ different $SO(N)$ representations in the product.
Note that the two seemingly degenerate antisymmetric tensor representations $L/R$ are in fact inequivalent\footnote{and map to each other under the parity operator $P\in O(N)$, such that the symmetric combination $[r]_L+[r]_R=[r]_{O(N)}$ and the anti--symmetric combination $[r]_L-[r]_R=[N-r]_{O(N)}$.} if we think of them as the representations of the $O(N)$ group. However, we will not distinguish between $[N/2]_\pm$. This gives us a total of $N+1$ representations for an $O(N)$ invariant density matrix. We can label these representations by
\begin{align}
    \left({\cal H}_L^\dagger \otimes {\cal H}_R \right)_{O(N)} = [0]+[1]+\cdots+[N]
\end{align}

In order to build an $O(N)$ invariant density matrix, we need to sum up all the states in the same representation with same weight.
Namely, the general density matrix can be written as a linear sum of projectors on each representation,
\begin{align} \label{eq:SO(N)DM}
    \rho=\sum_{i=0}^{N} a_i\rho_i
\end{align}
where $\rho_i$ denote the projectors on $N+1$ different representations of $O(N)$ included in the tensor product, while $a_i$ are arbitrary positive coefficients, such that $\Tr[\rho]=1$.

\paragraph{Basis of $O(N)$ invariant density matrices}  
Although $\rho_i$ comprise a set which spans the possible density matrices, there is more convenient option, which is similar to the $A\sigma\sigma$ that we used before. 
Consider the operator
\begin{align}
    Q=i\sum_{i=1}^N \chi_L^i\chi_R^i.
\end{align}
We will see that we can achieve any density matrix (\ref{eq:SO(N)DM}) by taking linear combinations of $Q^n$. 

First of all, $Q^2$ is related to a quadratic Casimir\footnote{The Pauli case $G(A)$'s can also be written using the quadratic casimir of $SU(2)$.} $\mathcal{C}_2$ of $SO(N)$. Recalling that the $SO(N)$ generators $\Sigma^{ij}$ are given by $\Sigma^{ij}\equiv -\frac{i}{2}\left([\chi^i_L,\chi^j_L]+[\chi^i_R,\chi^j_R]\right)$, then
\begin{align}
    \mathcal{C}_2\equiv \frac{1}{4}\sum_{i<j}({\Sigma^{ij}})^2=\frac{1}{2}\sum_{i<j}(1+\chi_L^i\chi_R^i\chi_L^j\chi_R^j)=\frac{1}{4}(N^2-Q^2).
\end{align}
For the antisymmetric representation of rank $0\leq r\leq N$, $\mathcal{C}_2$ is $\mathcal{C}_2=Nr-r^2$, which means
\begin{align}
    Q^2=\sum_{r} (N-2r)^2\rho_r.
\end{align}
One should also note that $Q^2$ is the same for rank $r$ and rank $N-r$ representations, and that $Q^2=0$ for the rank $N/2$ rep, irrespective of self-dual or anti-self-dual representations.

Now, since $Q$ itself is an $O(N)$ invariant operator, this can also be written as a sum of projection operators onto individual representations.
Also, since $Q$ has a single left fermion and a single right fermion we understand that it changes the chirality of the left and the right spinor at the same time. We see that $Q$ exchanges the two representations of the same dimensionality which were not distinguished by $Q^2$,
\begin{align}
\begin{split}
    [l]\in S_L\otimes S_R &\leftrightarrow  [l']\in {S'}_L\otimes {S'}_R\\
     [l]\in S_L\otimes {S'}_R &\leftrightarrow  [l']\in {S'}_L\otimes {S}_R
\end{split}
\end{align}
This means that there is a basis in which
\begin{align}
    Q=(N-2r)
    \begin{pmatrix}
    O & \mathbbm{1} \\
    \mathbbm{1} & O
    \end{pmatrix}
\end{align}
on a given representation of $SO(N)$, and upon diagonalization, $Q$ distinguishes the two representations of the same dimensionality with the sign of their eigenvalues.
We now conclude that
\begin{align}
    Q=\sum_{r=0}^{N} (N-2r)\rho_r.
\end{align}
We used the Hermiticity of $Q$ here.
This means that $Q^n$ spans the complete basis for the $O(N)$ invariant density matrix, for $0\leq n\leq N$, thus establishing our claim above.

\paragraph{Convenient basis of $O(N)$ invariant density matrices}

We have seen that the $O(N)$ invariant general density matrix is a function of $Q\equiv \sum_i\chi_L^i\chi_R^i$,
\begin{align}
    \rho=F\left[\sum_i\chi_L^i\chi_R^i\right].
\end{align}
Since $Q^{N+1}\subseteq \text{Sp}\{Q^0,\cdots,Q^N\}$, such a function is always of the form
\begin{align}
    \rho=F\left[\sum_i\chi_1^i\chi_2^i\right]=\sum_{i=0}^{N}a_i Q^i,
\end{align}
where $\Tr[\rho]=1$.
One can also take a more convenient basis
\begin{align}
    \rho=F\left[\sum_i\chi_L^i\chi_R^i\right]=\sum_{k=0}^{N} \frac{b_k\exp\left[i\alpha_kQ\right]}{\Tr\left[\exp\left[i\alpha_kQ\right]\right]}
\end{align}
where $\sum_{k=0}^{N}b_k=1$ to ensure $\Tr[\rho]=1$.

We can also rewrite this by using 
\begin{align}
    \frac{\exp\left[i\alpha_kQ\right]}{\Tr\left[\exp\left[i\alpha_kQ\right]\right]}
    =\prod_i \left[1+iA_k \chi_L^i\chi_R^i\right]\equiv \rho(A_k)
\end{align}
where $A_k=\tanh (\alpha_k)$, so that
\begin{align}
    \rho=F\left[\sum_i\chi_L^i\chi_R^i\right]=\sum_{k=0}^{N} b_k \rho(A_k)
\end{align}
The choice of $A_k$ is arbitrary, but it is usually convenient to take it so that 
$A_k=-1+\frac{2k}{N}$.
In the limit of $N\rightarrow\infty$, the basis of $O(N)$ invariant density matrices becomes $\rho(A)$, and we can take $A$ continuously from $-1$ to $1$.

\section{Crash course on chord diagrams}\label{sec:DSLimitAndCD}

In this appendix we will review the chord diagram techniques for the models introduced in \ref{sec:syk}, in the double scaling limit.

Consider the moments of the Hamiltonian (this is a re-iteration of (\ref{eq:Partition_Moment}))
\begin{align}
    {\hat m}_k = 2^{-N}m_k= 2^{-N} \left<\Tr(H^k)\right>_J. 
\end{align}
Let us now see how the double scaled limit allows us to compute it.

We will proceed with the random spin model as it will be useful for later discussion,  but the discussion is very similar in the Majorana SYK model, and performed in full in \cite{Berkooz:2018jqr}. We plug the Hamiltonian (\ref{eq:RandomSpinHamiltonian}) into (\ref{eq:Partition_Moment}), in order to get
\begin{align}
    {\hat m}_k = 3^{-kp/2}{\binom{N}{p}}^{-k/2}2^{-N}\sum_{I_1,\cdots,I_k}\left<I_1\cdots I_k\right>_J\Tr(\sigma_{I_1}\cdots\sigma_{I_k}).
\end{align}
Due to the Gaussian distribution (\ref{eq:Gaussian_Coeffs}), the expectation value over the coefficients is given by a sum over Wick contractions. This in turn means that the moment $m_k$ is given by all possible traces involving $k/2$ operator strings $\sigma_I$, each of which appears twice in $m_k$, as
\begin{align}
    m_k = 3^{-kp/2}{\binom{N}{p}}^{-k/2}2^{-N}\sum_{\text{Wick contractions}}\sum_{I_1,\cdots,I_{k/2}}\Tr(\sigma_{I_1}\sigma_{I_2}\cdots\sigma_{I_1}\cdots).
\end{align}
Each term in the sum over Wick contractions can be represented using a chord diagram. Let each of the $k$ operators $\sigma_I$ define a node on a circle. Each node is labelled by an index $j=1,\cdots,k$. We then connect the nodes in pairs, to designate which pairs have identical sets of sites $I_i$. See figure \ref{fig:SingleSided_CD_Example} for an example of a chord diagram. 

Now focus on a specific term in the double sum, i.e, a specific chord diagram (Wick contraction) with some specific choice of indices. We compute the trace by commuting the trace on each of the species. The obstruction to doing so is that some of the sites $i$ appear in more than a single index set $I$; If two chords do not intersect they contribute
\begin{align}
    \Tr(\sigma_{I_1}\sigma_{I_1}\sigma_{I_2}\sigma_{I_2}),
\end{align}
whereas if they intersect they give a factor proportional to 
\begin{align}
    \Tr(\sigma_{I_1}\sigma_{I_2}\sigma_{I_1}\sigma_{I_2}).
\end{align}
If there is a non-trivial overlap $I_1\cap I_2\neq\emptyset$ then these factors will be different.

In the double scaled limit described above, \cite{Erdos:2014zgc} showed that there are two major simplifications in computing this sum over chord diagrams:
\begin{enumerate}
    \item The number of overlapping indices between any two index sets $m_{ij}\equiv|I_i\cap I_j|$ is Poisson distributed with parameter $p^2/N$.
    \item With probability $1$, the intersection of any three index sets vanishes, namely $|I_i\cap I_j \cap I_k|=0$, for $i\neq j\neq k$. This statement is summarized in lemma $(9)$ there, and subsequent discussion.
\end{enumerate}
These two statements are just a consquence of the double scaling $p\propto \sqrt{N}$, and they hold in both the models described above\footnote{This is what determines the prefactors (\ref{eq:lambda}) for the two cases.}.

These simplifications imply the following. Consider now the sum over all index sets in some specific chord diagram, namely
\begin{align} \label{eq:beforeCD}
    3^{-kp/2}{\binom{N}{p}}^{-k/2}\sum_{I_1,\cdots,I_{k/2}}\Tr(\sigma_{I_1}\sigma_{J_2}\cdots\sigma_{J_1}\cdots),
\end{align} for some specific pairing, and start carrying out the traces in each of the species's Hilbert spaces. 

If a species appears only in a single chord, then the trace contributes ${\frac{1}{3}\cdot 3} \sum_a \Tr(\sigma^a\sigma^a)=1$ to ${\hat m}_k$.
Next, due to property (2) above, we can change the variables in the sum over $I_1,\cdots,I_{k/2}$ to the size of the overlap between pairs of index sets $m_{ij}=|I_i\cap I_j|$, with $i,j=1,\cdots,k/2$, along with the measure which is the probability of having a given overlap. Due to property (1) above, we know that this is exactly the Poisson distribution with parameter $p^2/N$. The ${\binom{N}{p}}^{-k/2}$ factor precisely turns counting of appearances of a certain type in the sum into probabilities of such events. Each overlap for a given intersection gives a factor of
\begin{align}
    3^{-2}\sum_{a,b=1}^{3}\frac{1}{2}\Tr\left(\sigma^{(a)}\sigma^{(b)}\sigma^{(a)}\sigma^{(b)}\right)=-\frac{1}{3},
\end{align}
relative to $1$ when the ordering is $aabb$, which originates from a pair of non-intersecting chords. 
Therefore, each intersection in the chord diagram gives a factor of
\begin{align} \label{eq:PoissonSum}
\sum_{m=0}^{\infty}\frac{(p^2/N)^m}{m!}e^{-p^2/N}(-1/3)^m=e^{-\lambda}= q.    
\end{align}
This factor is given for each chord intersection in a diagram. This allows us to bring the moment (\ref{eq:Partition_Moment}) to the final form
\begin{align} \label{eq:ChordPF}
    {\hat m}_k = \sum_{\text{CD}(k)}q^{\# \text{intersections}},
\end{align}
where $\text{CD}(k)$ represents chord diagrams with $k$ nodes.

\paragraph{Transfer matrix method} Next we will use some linear algebra in order to compute the weighted sum over all chord diagrams (\ref{eq:ChordPF}). Consider cutting the circle open at some point and going sequentially along the line. We define the Hilbert space $\mathcal{H}_{\text{aux}}$, which is spanned by $\{\ket{n}\}_{n=0}^{\infty}$, along with the diagonal inner product $\left<n|n'\right>=\delta_{nn'}$. We can think of $\left|n\right>$ as a state representing $n$ open chords, and a vector in the Hilbert space will be denoted by $\sum_{n\ge 0} v_n|n\rangle$. 

\begin{figure}
    \centering
    \includegraphics[width=0.5\textwidth,page=1]{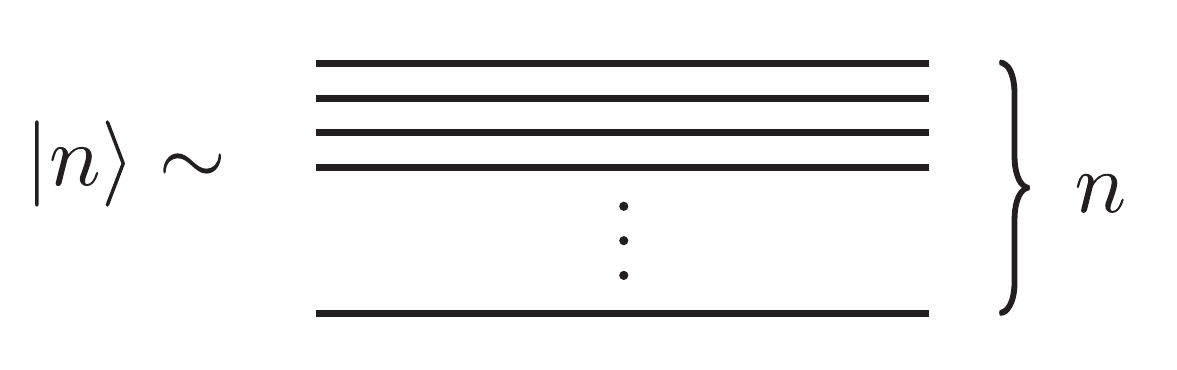}
    \caption{The vector $\ket{n}$ represents $n$ stacked chords.}
    \includegraphics[width=\textwidth,page=2]{Figures/TMatrixIntro.pdf}
    \caption{Acting with $T$}
    \label{fig:TMatrixExample}
\end{figure}

Define $T:\mathcal{H}_{\text{aux}}\to\mathcal{H}_{\text{aux}}$ the \textit{Transfer matrix} on $\mathcal{H}_{\text{aux}}$. We think of $T$ as acting on a state $\ket{n}$ by opening a new chord or closing an existing one, see figure \ref{fig:TMatrixExample}. We can reproduce the sum (\ref{eq:ChordPF}) if we decide that:
\begin{enumerate}
    \item $T$ always opens a new chord below all existing chords. This means that chords cannot intersect when they open.
    \item Whenever a chord closes and intersects another chord, it does so with a factor of $q$.
\end{enumerate}
This means that as we go over a node, the coefficients $v_n$ change by
\begin{align}
\begin{split}
    v_n(i+1) &=  v_{n-1}(i)+1\cdot v_{n+1}(i) +q\cdot  v_{n+1}(i) +\cdots+q^n\cdot v_{n+1}(i) \\
    & \qquad \qquad =v_{n-1}(i)+\frac{1-q^{n+1}}{1-q}v_{n+1}(i).
\end{split}
\end{align}
In this basis the matrix $T$ is given by 
\begin{align} \label{eq:TransferMatrix}
    T = \begin{pmatrix}
    0 & \frac{1-q}{1-q} & 0 & 0 & \cdots \\
    1 & 0 & \frac{1-q^2}{1-q} & 0 & \cdots \\
    0 & 1 & 0 & \frac{1-q^3}{1-q} & \cdots \\
    0 & 0 & 1 & 0 & \cdots \\
    \vdots & \vdots & \vdots & \vdots & \ddots
    \end{pmatrix}
\end{align}
Combining all of the above we see that in order to reproduce the sum appearing in (\ref{eq:ChordPF}) of all chord diagrams of length $k$, we need to consider the element 
\begin{align} \label{eq:TranMetMoment}
m_k = \left<0|T^k|0\right>.    
\end{align}

The task of finding the moment $m_k$ reduces to diagonalizing the matrix $T$ and taking its $k$'th power. This is done in \cite{Berkooz:2018qkz}, and we will not repeat the derivation here, but merely cite the results. We have
\begin{align}
    m_k = \int_0^\pi d\theta \frac{(q;q)_{\infty}|(e^{2i\theta};q)_{\infty}|^2}{2\pi}\cdot\left(\frac{2\cos\theta}{\sqrt{1-q}}\right)^k, 
\end{align}
where $(a;q)_n$ is the q-Pochammer symbol, defined by
\begin{align}
    (a;q)_n\equiv\prod_{k=0}^{n-1}(1-aq^k),
\end{align}
and when $n=\infty$ we extend the product to an infinite product.
By resumming the $m_k$ into the thermal partition function, we get
\begin{align}
    \Tr[e^{-\beta H}]=\int_0^\pi \frac{d\theta}{2\pi}(q,e^{\pm 2i\theta};q)_{\infty} \exp\left[-\beta \frac{2\cos\theta}{\sqrt{1-q}}\right],
\end{align}
where $(a_1,a_2,\dots,a_k;q)_n\equiv \prod_{i=1}^k (a_k;q)_n$, and $(e^{\pm i\theta};q)\equiv (e^{+ i\theta};q)(e^{- i\theta};q)$.
We refer the reader to \cite{Berkooz:2018jqr} for the computation of two and four-point functions.

\section{Special functions} \label{sec:SpecialFunctions}
In this section we assume $|q|<1$ and use the following variables:
\begin{align}
    \begin{split}
        x,y,x_i\in[-1,1], \qquad \theta,\phi,\theta_i\in[0,\pi],
        \\
        x=\cos(\theta), \quad y = \cos\phi,\quad x_i=\cos\theta_i.
        \end{split}
\end{align}
A $q$-Pochhammer symbol is defined as
\begin{align}
    (a;q)_n \equiv \prod_{k=1}^{n}\left(1-aq^{k-1}\right),
\end{align}
and we use a standard shorthand for their products
\begin{align}
    (a_1,a_2,\cdots;q)_n = (a_1;q)_n(a_2;q)_n\cdots.
\end{align}
In these formulas $n$ can be also set to infinity and the product converges.

A $q$-gamma function is defined via the $q$-Pochhammer symbol as
\begin{align}
    \Gamma_q(x) \equiv \frac{(q;q)_\infty}{(q^x;q)_\infty}(1-q)^x,\qquad 0<q<1.
\end{align}
The definition extends to $|q|<1$ by using principal values of $q^x$ and $(1-q)^{1-x}$. One can show that $\lim_{q\to1^{-}}\Gamma_q(x)=\Gamma(x)$.

We will use the standard shorthand for a basic one-variable well-poised hypergeometric series
\begin{align} \label{eq:DEF:8W7Series}
    _8W_7(a;b,c,d,e,f;q,z) \equiv \sum_{n=0}^{\infty}\frac{(a,\pm q a^{1/2},b,c,d,e,f;q)_n}{(\pm a^{1/2},qa/b,qa/c,qa/d,qa/e,qa/f,q;q)_n}z^n.
\end{align}
With an additional condition of very-well-poisedness
\begin{align}
    bcdefz=q^2a^2,
\end{align}
this function posses a $W(D_5)$ symmetry in its parameters, which is a bit hidden in any of the hypergeometric representations (\ref{eq:DEF:8W7Series}) (with just an $W(A_5)\equiv S_5$ part manifest). The additional symmetry generator corresponds to a so-called (limiting case of) Bailey transform \cite{gasper_rahman_2004}:
\begin{align}
    _8W_7\left(a;b,c,d,e,f;q,\frac{a^2q^2}{bcdef}\right) = \frac{\left(aq,\frac{aq}{ef},\frac{\lambda q}{e},\frac{\lambda q}{f};q\right)_\infty}{\left(\frac{aq}{e},\frac{aq}{f},\lambda q,\frac{\lambda q}{ef};q\right)_{\infty}}_8W_7\left(\lambda;\frac{\lambda b}{a},\frac{\lambda c}{a},\frac{\lambda d}{a},e,f;q,\frac{aq}{ef}\right),
\end{align}
where $\lambda \equiv qa^2/bcd$ and we require
\begin{align}
    \left|\frac{aq}{ef}\right|<1,\quad \left|\frac{\lambda q}{ef}\right|<1
\end{align}
for convergence.

The following Mellin-Barnes-Agarwal integral representation for very-well-poised $_8W_7$ is known \cite{gasper_rahman_2004}
\begin{align} \label{eq:W_Integral_rep}
    \begin{split}
        &_8W_7(q^A;q^a,q^b,q^c,q^d,q^e;q,q^B) = \sin[\pi(a+b+c-A)]
        \\
        &\quad \times \frac{(q^{1+A},q^a,q^b,q^c,q^{1+A-a-b},q^{1+A-b-c},q^{1+A-a-c},q^{1+A-d-e};q)_\infty}{(q,q^{a+b+c-A},q^{1+A-a-b-c},q^{1+A-a},q^{1+A-b},q^{1+A-c},q^{1+A-d},q^{1+A-e};q)_\infty}
        \\
        & \quad \times \int_{-i\infty}^{i\infty}\frac{ds}{2\pi i}\frac{\pi q^s}{\sin (\pi s)\sin (\pi[a+b+c-A+s])}\frac{(q^{1+s},q^{1+A-d+s},q^{1+A-e+s},q^{a+b+c-A+s};q)_\infty}{(q^{a+s},q^{b+s},q^{c+s},q^{1+A-d-e+s};q)_{\infty}}
    \end{split}
\end{align}
for the parameters satisfying $B=2+2A-a-b-c-d-e$, such that
\begin{align}
    \text{Re}B>0,\qquad \text{Re}(s\log q-\log(\sin\pi s\sin\pi (a+b+c-A+s))).
\end{align}
If the last condition is not satisfied on the entire imaginary line, the contour should be indented according to a usual Mellin-Barnes prescription (i.e. separating poles going to right from going to the left). When phrased in terms of gamma and $q$-gamma functions, the above integral representation is immediately seen to reduce in $q\to 1^-$ limit to a Mellin-Barnes representation of the corresponding (undeformed)  well-poised hypergeometric $_7F_6(1)$, i.e. a Wilson function \cite{2005math......1511G} (up to appropriate Pochhammer factors).

\subsection{R-matrix, integral form} \label{sec:R_Matrix_integral_form}

Here we would like to obtain a convenient integral form for (\ref{eq:DoubleSided2pt}). First, we rewrite it as
\begin{align}
\begin{split}
  &\int_0^{\pi}\prod_{j=1}^{4}\left\{\frac{d\theta_j}{2\pi}(q,e^{\pm2i\theta_j};q)_{\infty}\right\}e^{-\beta_1(E(\theta_4)+E(\theta_3)-\beta_2(E(\theta_2)+E(\theta_1))}  
  \\
  &\qquad  \times \left(\frac{(B^2;q)_{\infty}}{(Be^{i(\pm\theta_1\pm\theta_4)};q)_{\infty}}\frac{(B^2;q)_{\infty}}{(Be^{i(\pm\theta_2\pm\theta_3)};q)_{\infty}}\frac{(\tilde{q}^2;q)_{\infty}}{(\tilde{q}e^{i(\pm\theta_1\pm\theta_2)};q)_{\infty}}
  \frac{(\tilde{q}^2;q)_{\infty}}{(\tilde{q}e^{i(\pm\theta_3\pm\theta_4)};q)_{\infty}}\right)^{1/2}R^{(q)}_{\theta_4\theta_2}\begin{bmatrix}
  \theta_3 & \ell_M \\
  \theta_1 & \ell_B
  \end{bmatrix},
\end{split}
\end{align}
with 
\begin{align}
\begin{split}
        R^{(q)}_{\theta_4\theta_2}\begin{bmatrix}
  \theta_3 & \ell_M \\
  \theta_1 & \ell_B
  \end{bmatrix} &= \frac{(Be^{-i(\theta_2+\theta_3)},B\tilde{q}e^{i(\theta_3\pm\theta_1)},B\tilde{q}e^{i(\theta_2\pm\theta_4)};q)_{\infty}}{(B\tilde{q}^2e^{i(\theta_2+\theta_3)};q)_{\infty}}
  \\
  & \qquad \times \frac{(\tilde{q}^2;q)_{\infty}}{[(Be^{i(\pm\theta_2\pm\theta_3)},Be^{i(\pm\theta_1\pm\theta_4)},\tilde{q}e^{i(\pm\theta_1\pm\theta_2),\tilde{q}e^{i(\pm\theta_3\pm\theta_4)}}]^{1/2}}
  \\
  & \qquad \times _8W_7\left(\frac{B\tilde{q}^2e^{i(\theta_2+\theta_3)}}{q};Be^{i(\theta_2+\theta_3)},\tilde{q}e^{i(\theta_2\pm\theta_1)},\tilde{q}e^{i(\theta_3\pm\theta_4)};q,Be^{-i(\theta_2+\theta_3)}\right).
\end{split}
\end{align}
Let us now use the In section 6 of \cite{Berkooz:2018jqr}, it is shown how to use the symmetries of the $_8W_7$ function, as well as the integral representation (\ref{eq:W_Integral_rep}) in order to write the R-matrix as
\begin{align}
\begin{split}
   & R^{(q)}_{\theta_4\theta_2}\begin{bmatrix}
  \theta_3 & \ell_M \\
  \theta_1 & \ell_B
  \end{bmatrix} \\&= -\frac{(1-q)^{-2}}{\Gamma(1+\ell_M-\ell_B+iy_1-iy_3)\Gamma(\ell_B-\ell_M+iy_3-iy_1)}\frac{1}{\left(q,\frac{q\tilde{q}}{B}e^{i(\theta_1-\theta_3)},\frac{B}{\tilde{q}}e^{i(\theta_3-\theta_1)};q\right)_{\infty}}
  \\
  &\times \sqrt{\frac{(Be^{i(\theta_3\pm\theta_2)},Be^{i(-\theta_1\pm\theta_4)},\tilde{q}e^{i(\theta_1\pm\theta_2)},\tilde{q}e^{i(-\theta_3\pm\theta_4)})_{\infty}}
  {(Be^{i(-\theta_3\pm\theta_2)},Be^{i(\theta_1\pm\theta_4)},\tilde{q}e^{i(-\theta_1\pm\theta_2)},\tilde{q}e^{i(\theta_3\pm\theta_4)})_{\infty}}}
  \\
  & \times \int_{\mathcal{C}}\frac{ds}{2\pi i}q^s\Gamma(1+s)\Gamma(-s)\Gamma(s+1-\ell_B+\ell_M+iy_1-iy_3)\Gamma(-s+\ell_B-\ell_M-iy_1+iy_2)
  \\
  & \times \frac{\Gamma_q(s+\ell_M+iy_1-iy_2)\Gamma_q(s+\ell_M-iy_3\pm iy_4)\Gamma_q(s+\ell_M+iy_1+iy_2)}{\Gamma_q(s+1)\Gamma_q(s+\ell_B+\ell_M+iy_1-iy_3)\Gamma_q(s+2\ell_M)\Gamma_q(s+1+\ell_M-\ell_B+iy_1-iy_3)},
 \end{split}
\end{align}
where the $y_i$ are defined in (\ref{eq:Def:y_var}). Notice we use these variables even though we don't necessarily restrict ourselves to the low energy limit. The contour $\mathcal{C}$ is a deformation of the contour going along the imaginary axis, such that the poles that come from Gamma functions with $(+s)$ arguments are to the left of $\mathcal{C}$, and those that come from Gamma $(-s)$ are to the right of it (a usual Mellin-Barnes prescription). Then we shift the integration variable $s\to s-\ell+iy_3-iy_4$ (with no other contributions because of the Mellin-Barnes prescription) and express all the $q$-Pochhammers in terms of $\Gamma_q$, to get
\begin{align}
    \begin{split}
         &R^{(q)}_{\theta_4\theta_2}\begin{bmatrix}
  \theta_3 & \ell_M \\
  \theta_1 & \ell_B
  \end{bmatrix}
  = -\frac{1}{(q;q)^3_{\infty}(1-q)^3}\cdot\frac{\Gamma_q(1+\ell_M-\ell_B+iy_1-iy_3)\Gamma_q(\ell_B-\ell_M+iy_3-iy_1)}{\Gamma(1+\ell_M-\ell_B+iy_1-iy_3)\Gamma(\ell_B-\ell_M+iy_3-iy_1)}
  \\
  & \sqrt{\frac{\Gamma_q(\ell_B-iy_3\pm iy_2)\Gamma_q(\ell_B+iy_1\pm iy_4)\Gamma_q(\ell_M-iy_1\pm iy_2)\Gamma_q(\ell_M+iy_3\pm iy_4)}
  {\Gamma_q(\ell_B+iy_3\pm iy_2)\Gamma_q(\ell_B-iy_1\pm iy_4)\Gamma_q(\ell_M+iy_1\pm iy_2)\Gamma_q(\ell_M-iy_3\pm iy_4)}}
  \\
  & \int_{\mathcal{C}}\frac{ds}{2\pi i}q^{s-\ell_M+iy_3-iy_4}\frac{\Gamma(s+1-\ell_M+iy_3-iy_4)\Gamma(s+1-\ell_B+iy_1-iy_4)}
  {\Gamma_q(s+1-\ell_M+iy_3-iy_4)\Gamma_q(s+1-\ell_B+iy_1-iy_4)}
  \\
  &\frac{\Gamma_q(s)\Gamma_q(s-2iy_4)\Gamma_q(s+iy_1+iy_3-iy_4\pm iy_2)\Gamma(-s+\ell_M-iy_3+iy_4)\Gamma(-s+\ell_B+iy_4-iy_1)}{\Gamma_q(s+\ell_B+iy_1-iy_4)\Gamma_q(s+\ell_M+iy_3-iy_4)}.
    \end{split}
\end{align}

\bibliographystyle{JHEP}
\bibliography{EntBib}

\end{document}